\documentclass[iop,numberedappendix,appendixfloats]{emulateapj}
\usepackage{epsfig}
\usepackage[usenames,dvips]{color}
\usepackage{bibcontents}
\usepackage{graphicx}
\usepackage{amsmath}
\usepackage{natbib}
\usepackage{amssymb}
\usepackage{amstext}
\usepackage[margin=20pt,font=small,labelfont=bf]{caption}
\usepackage{subfig}
\usepackage{multirow}
\usepackage{array}
\usepackage{placeins}
\usepackage{booktabs}
\usepackage{varwidth} %for the varwidth minipage environment, for table
\usepackage{verbatim}

\begin{document}

\title{Are the Effects of Structure Formation Seen in the Central Metallicity \\ of Galaxy Clusters?}
\shorttitle{Metallicity and Structure Formation}

\author{
Tamer Y. Elkholy\altaffilmark{1,2},
Mark W. Bautz\altaffilmark{2},
Claude R. Canizares\altaffilmark{1,2}
}

\altaffiltext{1}{Department of Physics, Massachusetts Institute of Technology, 77 Massachusetts Avenue, Cambridge, Massachusetts 02139, USA; \email{tykholy@gmail.com}}
\altaffiltext{2}{MIT Kavli Institute for Astrophysics and Space Research, 77 Massachusetts Avenue, Cambridge, Massachusetts 02139, USA}
\date{\today}

\begin{abstract}

A sample of 46 nearby clusters observed with \textit{Chandra} is analyzed to produce radial density, temperature, entropy and metallicity profiles, as well as other morphological measurements. The entropy profiles are computed to larger radial extents than in previous \textit{Chandra} cluster sample analyses. We find that the iron mass fraction measured in the inner $0.15R_{500}$ shows a larger dispersion across the sample of low-mass clusters, than it does for the sample of high-mass clusters. We interpret this finding as the result of the mixing of more haloes in large clusters than in small clusters, which leads to an averaging of the metal content in the large clusters, and thus less dispersion of metallicity for high-mass clusters. This interpretation lends support to the idea that the low-entropy, metal-rich gas of merging haloes reaches clusters' centers, which explains observations of Core-Collapse Supernova products metallicity peaks, and which is seen in hydrodynamical simulations. The gas in these merging haloes would have to reach the centers of clusters without mixing in the outer regions, in order to support our interpretation. On the other hand, metallicity dispersion does not change with mass in the outer regions of clusters, suggesting that most of the outer metals come from a source with a more uniform metallicity level, such as during pre-enrichment. We also measure a correlation between the metal content in low-mass clusters and the degree to which their Intra-Cluster Medium (ICM) is morphologically disturbed, as measured by centroid shift. This suggests an alternative interpretation of the large width of the metallicity distribution in low-mass clusters, whereby a metallicity boost in the center of low-mass clusters is induced as a transitional state, during mergers.

\end{abstract}


\section{Introduction}

In today's Universe, galaxy clusters are the largest bound structures, ranging in mass from about $10^{13}$ to $10^{15}M_\odot$. Their formation is understood within the paradigm of hierarchical structure formation, where large structures form through the merger of smaller structures. Because of their large masses, the Intra-Cluster Medium (ICM) in galaxy clusters is heated and compressed, such that bremsstrahlung emission becomes an efficient radiation process. This X-ray emission from the ICM contains within it ample information about the chemistry and the dynamical state of the ICM, and by extension the cluster as a whole. In particular we focus in this study on two quantities, metallicity and entropy, because they encode the integrated effect of various physical processes occurring in clusters.

Most metals have been produced in stars, predominantly through thermonuclear burning, or in processes resulting from the extreme conditions in supernovae (SNe). Therefore, studying metallicity in galaxy clusters probes processes that produce metals, such as star formation and SN rates, as well as processes that mix and distribute metals such as mergers and central AGN activity.

\citet{renzi93} were the first to point out that one cannot account for all the observed metals in the ICM assuming the current SN Type Ia rate per unit luminosity was the same in the past. This problem was studied with better SN data and more complex chemical evolution and ICM enrichment models by \citet{porti04} and \citet{loewe06}, who both showed the need for more metal production in clusters, compared to what is observed in the field. Both studies above in addition to \citet{nagas05}, suggest that a different stellar Initial Mass Function (IMF) must operate in field galaxies than in clusters. For example, a top-heavy IMF might be necessary in galaxy clusters.

Hydrodynamical models predicting the metallicity of the ICM have been reviewed by \citet{borga08}. The results of these studies in terms of the effect of different IMF's on the level of enrichment vary, as some are able to match or even exceed the observed metallicity in clusters, while others cannot match it, depending on the many details and parameters of the models. 

Two classes of SNe are responsible for producing the metals in clusters. Core Collapse SNe (SNCC) result from massive stars, whose lives are short compared to the time since the peak of star formation, at a redshift $z \sim 2-3$. Type Ia SNe (SNIa) result from low-mass stars and thus can be delayed by billions of years from the time of star formation. If the above picture of a prompt metal injection from SNCC, followed by a more gradual injection from SNIa holds, then we should expect this to affect the metallicity radial profiles. Each class of SNe produces a different set of elements \citep[e.g.][]{werne08}. Both SN classes produce iron, while, for example, SNCC are the main contributors of oxygen, neon and magnesium. Since the rate of SNCC at any epoch is roughly proportional to the then-current rate of star formation, and since clusters form relatively late,  we expect SNCC products to have enriched the proto-ICM before cluster collapse. The SNCC products are therefore expected to be homogeneously spread through the ICM. Conversely, SNIa products are expected to be nearer the center of the cluster, as the Brightest Cluster Galaxy (BCG) accumulates more and more stars with time. However, using very deep \textit{Chandra} observations, \citet{sande06}, \citet{simio09}, \citet{simio10} and \citet{milli11} measure a large amount of SNCC products in the core of nearby galaxy clusters, presenting a challenge to the above picture.

Because entropy remains unchanged under adiabatic processes, it measures any heat input or output to the gas. In the case of the ICM, the main process that changes its entropy is gravitational shock heating. Other processes affecting the ICM entropy include radiative cooling, heating by AGN and SNe, as well as processes such as turbulent dissipation and conduction. The reader is referred to \citet{mcnam07} for a detailed review of the above processes.

Gravity is expected to be the dominant force in cluster formation, and thus most of the entropy generation is expected to result from shock heating of infalling gas, as a cluster forms. This process converts gravitational potential energy into heat. Because $S \propto T/n^{2/3}$, we expect the characteristic entropy scale of a cluster to scale proportionally to its virial temperature, or as $M^{2/3}$, where $M$ is the cluster mass. Hydrodynamical simulations \citep[e.g.][]{voit05,nagai07} produce a normalization of the entropy expected from gravity, as well as a radial profile, which is proportional to a power-law of radius, $r^{1.1-1.2}$. Observationally, \citet{ponma99} were among the first to measure the departure of the entropy from the expected self-similar scaling with cluster mass. The same result was later found in \citet{lloyd00}, \citet{ponma03}, and \citet{pratt10}.

Some studies have focused on cases where the effects of both heating and metal enrichment can be detected. High-resolution observations of the centers of clusters find regions of enhanced metallicity expected to have been ejected from the central regions of clusters, by the central engine \citep[e.g.][]{simio09,simio09b,kirk09,osull11}. For example, \citet{kirk11} showed that the direction of elongation of an enhanced-metallicity region is correlated with the direction of cavities and radio emission axes, originating from the center. On larger scales, however, simulations by \citet{borga05} and \citet{short13} suggest that winds from SNe have little effect on the overall entropy of clusters. There is, however, a shortage of attempts to test these results observationally.

The aim of this work is therefore to apply a systematic, spatially resolved study of metallicity and entropy to a large sample of galaxy clusters, to look for any possible relation between the metallicity and the entropy of the ICM. We also aim to produce a dataset of entropy profiles (in the form of temperature and metallicity profiles) for the community to employ in various galaxy cluster studies, and make it available electronically.

Throughout this work a $\Lambda$CDM cosmological model is assumed, where the Hubble constant is $H_0=72$ km/s/Mpc, the matter density parameter $\Omega_m=0.26$, the dark energy density is $\Omega_\Lambda=0.74$, and the universal baryon fraction, $f_b=\Omega_b/\Omega_m=0.169$, where $\Omega_b$ is the baryon density.

\section{Data Sample and Analysis}

\subsection{Data Sample}
Our sample consists of bright clusters --- present in both the HIFLUGCS and ACCEPT samples --- which were observed with \textit{Chandra}'s ACIS instrument out to at least $0.2R_{500}$, where $R_{500}$ is the radius enclosing an average density that is 500 times the critical density of the Universe at the redshift of the observed cluster. More precisely, we start with the extended HIFLUGCS sample of galaxy clusters, which is a flux-limited sample of clusters with X-ray flux $f_X[0.1-2.4\mbox{k}eV] \geq 2 \times 10^{-11}$ erg/s/cm$^2$ \citep{reipr02}. We search for the HIFLUGCS clusters which have \textit{Chandra} coverage\footnote{For the purpose of selecting clusters, we use the values of $R_{500}$ measured in \citet{reipr02}.} out to at least $0.2R_{500}$. Once the above sample is identified, we search the ACCEPT data for the aforementioned HIFLUGCS subsample, where entropy is measured in at least 3 radial bins beyond $0.2R_{500}$. The ACCEPT study \citep{cavag09} measured the entropy profiles of all galaxy clusters observed by \textit{Chandra}, up to August 2008, and has made the data available online\footnote{http://www.pa.msu.edu/astro/MC2/accept/}. Forty eight galaxy clusters satisfy the above selection criteria.

For all spectral analysis, we discard all observations taken before January 29, 2000 as the \textit{Chandra} focal plane temperature for that period was $-110^{\circ}$C or higher, which increases the level of background. \textit{Chandra}'s calibration is in general better for later dates. Only one cluster, Abell 401, is excluded from our analysis because it was only observed prior to January 29, 2000. We also exclude Abell 2255 because its available observation was too short to yield enough photons. The complete analysis will thus be presented for a sample of 46 clusters.

We show in Table \ref{tab:datasample} the observational details of our sample, and the \textit{Chandra} observations used in this work. For each cluster, we show the label which will be used to designate it hereafter, in figures and tables. We show the cluster's celestial coordinates, and its redshift. We list the \textit{Chandra} Observation ID's (OBSID) we use in this work. Some OBSID's are shown in parentheses. These are the observations that were used in imaging analysis, and excluded from spectral analysis, as described above. Finally, we show the total exposure time used in the spectral analysis of each cluster.

\newcolumntype{M}{>{\begin{varwidth}{4cm}}l<{\end{varwidth}}}
\newcolumntype{N}{>{\begin{varwidth}{1cm}}l<{\end{varwidth}}}

\begin{table*}[h] 
\begin{center}
\caption{\textit{Chandra} Observations.\label{tab:datasample}}
\begin{small} 
\begin{tabular}{ @{}lNlllMl@{} } 
% table made by fromoffice/makedatatable_postref.py
\toprule
Cluster\tablenotemark{a} & Label\tablenotemark{b} & RA\tablenotemark{c} & Dec\tablenotemark{d} & $z$\tablenotemark{e} & Obsids\tablenotemark{f} & $T$ (ks)\tablenotemark{g} \\
\midrule 
Abell 119 & a119 & 00:56:15.392 & -01:15:17.78 & 0.044 & 4180, 7918 & 57 \\
Abell 1413 & a1413 & 11:55:17.986 & +23:24:17.82 & 0.1427 & 1661, 5002, 5003 & 121 \\
Abell 1644 & a1644 & 12:57:11.772 & -17:24:33.68 & 0.0474 & 2206, 7922 & 70 \\
Abell 1651 & a1651 & 12:59:22.188 & -04:11:45.80 & 0.086 & 4185 & 10 \\
Abell 1689 & a1689 & 13:11:29.495 & -01:20:29.02 & 0.184 & 5004, 540, 6930, 7289 & 181 \\
Abell 1736 & a1736 & 13:26:54.235 & -27:09:48.65 & 0.0461 & 10428, 10429, 10430, 10431, 4186 & 35 \\
Abell 1795 & a1795 & 13:48:52.668 & +26:35:30.73 & 0.0616 & 10898, 10899, 10900, 10901, 12026, 12028, 12029, 13106, 13107, 13108, 13109, 13110, 13111, 13112, 13113, 13412, 13413, 13414, 13415, 13416, 13417, 5286, 5287, 5288, 6159, 6160, 6161, 6162, 6163, (494) & 437 \\
Abell 1914 & a1914 & 14:26:01.072 & +37:49:32.97 & 0.1712 & 3593, (542) & 27 \\
Abell 2029 & a2029 & 15:10:56.091 & +05:44:40.94 & 0.0767 & 10437, 4977, 6101, 891 & 112 \\
Abell 2063 & a2063 & 15:23:05.323 & +08:36:28.49 & 0.0354 & 5795, 6263 & 27 \\
Abell 2065 & a2065 & 15:22:29.060 & +27:42:34.33 & 0.0721 & 3182 & 22 \\
Abell 2142 & a2142 & 15:58:20.103 & +27:13:58.76 & 0.0899 & 5005, (1196, 1228) & 68 \\
Abell 2147 & a2147 & 16:02:15.608 & +15:57:53.77 & 0.0351 & 3211 & 18 \\
Abell 2163 & a2163 & 16:15:46.519 & -06:08:50.57 & 0.201 & 1653, 2455, 545 & 89 \\
Abell 2204 & a2204 & 16:32:46.922 & +05:34:31.40 & 0.1523 & 499, 6104, 7940 & 97 \\
Abell 2244 & a2244 & 17:02:42.517 & +34:03:37.46 & 0.097 & 4179 & 57 \\
Abell 2256 & a2256 & 17:03:59.388 & +78:38:44.57 & 0.0601 & 2419, (1386, 965) & 35 \\
Abell 2319 & a2319 & 19:21:09.997 & +43:57:18.82 & 0.0564 & 3231 & 14 \\
Abell 2657 & a2657 & 23:44:56.531 & +09:11:28.75 & 0.0404 & 4941 & 16 \\
Abell 2734 & a2734 & 00:11:21.616 & -28:51:17.98 & 0.062 & 5797 & 20 \\
Abell 3112 & a3112 & 03:17:57.627 & -44:14:20.34 & 0.075 & 2216, 2516, 6972, 7323, 7324 & 108 \\
Abell 3158 & a3158 & 03:42:52.591 & -53:37:50.03 & 0.059 & 3201, 3712 & 56 \\
Abell 3376 & a3376 & 06:01:57.312 & -39:58:25.80 & 0.0455 & 3202, 3450 & 64 \\
Abell 3391 & a3391 & 06:26:20.780 & -53:41:32.98 & 0.0531 & 4943 & 18 \\
Abell 3571 & a3571 & 13:47:28.580 & -32:51:14.35 & 0.0397 & 4203 & 34 \\
Abell 3667 & a3667 & 20:12:36.316 & -56:50:40.74 & 0.056 & 5751, 5752, 6292, 6295, 6296, 889, (513) & 430 \\
Abell 3822 & a3822 & 21:54:06.292 & -57:51:41.06 & 0.076 & 8269 & 8 \\
Abell 3827 & a3827 & 22:01:53.279 & -59:56:45.99 & 0.098 & 7920 & 46 \\
Abell 3921 & a3921 & 22:49:57.845 & -64:25:44.13 & 0.0936 & 4973 & 29 \\
Abell 399 & a399 & 02:57:51.557 & +13:02:32.43 & 0.0715 & 3230 & 49 \\
Abell 400 & a400 & 02:57:41.119 & +06:01:20.03 & 0.024 & 4181 & 21 \\
Abell 4038 & a4038 & 23:47:43.200 & -28:08:38.30 & 0.0283 & 4188, 4992 & 40 \\
Abell 4059 & a4059 & 23:57:00.933 & -34:45:34.44 & 0.046 & 5785, 897 & 110 \\
Abell 478 & a478 & 04:13:25.199 & +10:27:53.90 & 0.09 & 1669, 6102 & 52 \\
Abell 539 & a539 & 05:16:36.680 & +06:26:34.63 & 0.0288 & 5808, 7209 & 43 \\
Abell 644 & a644 & 08:17:25.392 & -07:30:48.38 & 0.0704 & 10420, 10421, 10422, 10423, 2211 & 49 \\
Abell 754 & a754 & 09:09:21.084 & -09:41:05.78 & 0.0528 & 10743, 6793, 6794, 6796, 6797, 6799, (577) & 187 \\
Abell S 405 & as405 & 03:51:29.787 & -82:13:21.26 & 0.0613 & 8272 & 8 \\
Hydra A & hyda & 09:18:05.876 & -12:05:43.17 & 0.0538 & 4969, 4970, (575, 576) & 239 \\
Zw III 54 & iiizw54 & 03:41:17.508 & +15:23:54.82 & 0.0311 & 4182 & 23 \\
MKW 3S & mkw3s & 15:21:51.708 & +07:42:24.65 & 0.045 & 900 & 57 \\
MKW 8 & mkw8 & 14:40:39.353 & +03:28:03.08 & 0.027 & 4942 & 23 \\
PKS 0745-191 & pks0745-191 & 07:47:31.265 & -19:17:41.62 & 0.1028 & 2427, 508, 6103 & 56 \\
UGC 3957 & ugc3957 & 07:40:58.133 & +55:25:38.25 & 0.034 & 8265 & 8 \\
ZwCl 1215+0400 & z1215 & 12:17:41.934 & +03:39:39.74 & 0.075 & 4184 & 12 \\
ZwCl 1742+3306 & z1742 & 17:44:14.447 & +32:59:29.02 & 0.0757 & 11708, 8267 & 53 \\
\bottomrule 
\end{tabular} 
\tablenotetext{1}{Cluster name.}
\tablenotetext{2}{Label used to denote cluster.}
\tablenotetext{3}{Cluster right ascention.}
\tablenotetext{4}{Cluster declination.}
\tablenotetext{5}{Cluster redshift.}
\tablenotetext{6}{The $Chandra$ OBSID's used in this work. OBSID's in parantheses were used in imaging analysis, and excluded from spectral analysis.}
\tablenotetext{7}{The total exposure time for spectral analysis in kiloseconds.}
\end{small} 
\end{center} 
\end{table*}

\subsection{Data Preparation}
\label{sec:dataprep}
We use the Chandra Interactive Analysis of Observations software, more commonly known as CIAO\footnote{http://cxc.harvard.edu/ciao/}, for analysis in this work. More precisely, we use CIAO's 4.2 version. Data are reprocessed following the guidelines in the CIAO analysis thread ``Reprocessing Data to Create New Level=2 Event File", using the tool \texttt{acis\_process\_events}. This reprocessing includes filtering to keep only event grades 0, 2, 3, 4 and 6. In addition, the VFAINT background cleaning method is applied to observations in VFAINT mode, using the \texttt{check\_vf\_pha=yes} option to \texttt{acis\_process\_events}.

The center of the clusters is defined to be the centroid of event x- and y-coordinates, calculated using the following iterative scheme. First, we select the observation with the longest exposure time, in the cases where we have multiple observations of one cluster. For ACIS-I pointing, we calculate the centroid using all four ACIS-I chips, while for ACIS-S pointings, we only use the chip with the most counts. We only include events with energies between 0.3 and 7keV. For the first centroid computation iteration, Iteration 1, we calculate the medians and the standard deviations of the x- and y-coordinates of the events from the entirety of the selected chip(s) of the longest-exposure observation. For Iteration 2, we restrict the median and standard deviation calculation to events within an ellipse with semi-major axes equal to \textit{twice} the x and y standard deviations calculated in Iteration 1, i.e. $2 \times \left( \sigma_{x1},\sigma_{y1} \right)$. For Iteration 3, the ellipse is shrunk to have semi-major axes equal to $1 \times \left( \sigma_{x2},\sigma_{y2} \right)$. Similarly, Iteration 4's filter ellipse has semi-major axes $0.75 \times \left( \sigma_{x3},\sigma_{y3} \right)$. For the final iteration, Iteration 5, we employ events from all observations in the centroid computation, instead of using the longest-exposure observation alone. The Iteration 5 filter ellipse has semi-major axes equal to $1 \times \left( \sigma_{x4},\sigma_{y4} \right)$.

Events from point sources are then identified and discarded. We use CIAO's \texttt{wavdetect} tool, applied to an image of the merged event files from all observations. The input image to \texttt{wavdetect} includes only events with energies in the range 0.3-7keV and is binned in $\left(2 \times 2 \right)$-pixel bins. The detected point source regions are inspected by eye to ensure that each region is large enough to include all events from its corresponding detected point source, and to add sources that were not detected by \texttt{wavdetect}. The latter tend to be point sources away from the telescope's optical axis, where the point spread function is much larger than it is in the center. We also exclude any region of bright extended emission, which does not belong to the central cluster emission, such as that from infalling sub-clusters (e.g. the sub-cluster to the North of Abell 2163.) We expect that many clusters will contain emission from faint infalling sub-clusters, which cannot be resolved due to their low surface brightness. Therefore we do not attempt to discard all emission from identified sub-clusters, and only remove the bright peaks of such emission when present.

Periods of high count rates resulting from flares are removed using the \texttt{lc\_clean()} tool in \textit{Sherpa}, which is CIAO's tool for spectral analysis. We compute a light curve of all data counts, excluding the point sources detected above, and the central 300'' to exclude the bulk of the cluster emission. Short flares are excluded when they are identified by \texttt{lc\_clean()}, while longer flares are excluded manually by selecting events in the time range which is sufficiently removed from the the start or the end of the flare. Some observations in which one or many flares last for most of the exposure time are entirely excluded.

For each observation, we create a background dataset from the Blank Sky files, available as part of the \textit{Chandra} calibration files. We choose the Blank Sky file for each ACIS chip based on the cluster dataset's observation date, its aim point and whether a CTI correction was applied to it. As mentioned above, we exclude all data taken prior to January 29, 2000. No Blank-Sky datasets are available in Period C, in VFAINT mode. We therefore assign to these datasets, the Blank-Sky files from Period D in VFAINT mode. In addition, we exclude some ACIS-S data taken during Period C, because their corresponding Blank-Sky files are not available\footnote{The definition of the background periods used here is available in Maxim Markevitch's note at http://cxc.harvard.edu/contrib/maxim/acisbg/COOKBOOK}.

\subsection{Data Analysis}

As is custom in clusters astrophysics, we define entropy as $S = kT/n_e^{2/3} $, where $k$ is Boltzmann's constant, $T$ the ICM temperature, and $n_e$ its electron density. As defined above, $S$ is related to the thermodynamical entropy per particle, $s$, through $s = (3k/2) \ln S + s_0$, where $s_0$ only depends on fundamental constants. We assume spherical symmetry and compute the entropy radial profiles from the density and temperature profiles, as described below.

\subsubsection{Density Profile, $n_e(r)$}
\label{sec:nofr}

Computing the electron density profile of the ICM is done in two steps. First, we use surface brightness (SB) measurements to constrain the shape of the density radial profile. Then, we use spectral measurements to set the overall normalization of $n_e(r)$. To fit for the density profile shape, we extract a SB profile based on photon counts in the energy range 0.7--2keV. Radial bins are defined such that boundary radii are spaced logarithmically, with a constant ratio of 1.25 between neighboring radii, except when this spacing results in fewer than 100 counts in the bin, in which case it is extended to the next radius. The innermost radius is defined as the projection of 1.2 arcseconds in the plane of the cluster, in order to include the largest number of counts, while avoiding any potential point source coinciding with the centroid of the ICM X-ray emission. From the source counts in each radial bin, we subtract the background contribution computed from the blank-sky datasets, and scaled by detector area and exposure time to match each observation and radial bin. The net number of counts is then normalized by the exposure map, to correct for the position dependence of \textit{Chandra}'s effective area. Finally we obtain a count SB by dividing by the solid angle of the extraction region.

The emissivity of an X-ray plasma at cluster temperatures is primarily in the form of bremsstrahlung and line emission. The contribution of each of those processes is proportional to $n_e^2$, but depends differently on the plasma temperature. However, when emissivity is integrated over the 0.7--2keV energy range, the different temperature dependencies contrive to cancel each other, and the resulting emission in this energy band has a negligible temperature dependence. In addition, we assume the ICM to be optically thin. Therefore, the SB at any given point on the sky is simply the integral of all emission along the line of sight to that point. We perform a maximum-likelihood fit to find the density profile that best fits the counts SB profile, up to a normalization. We use an analytical form for $n_e(r)$ that is flexible enough to allow us to phenomenologically fit the SB in all radial bins. Namely, we choose a simplified version of Vikhlinin's extended beta model \citep{vikh06},

\begin{equation}
\label{eq:vikhnofr}
\frac{n_e^2(r)}{n_0^2}  =  \frac {(\frac{r}{r_c})^{-\alpha}} {(1+\frac{r^2}{r_c^2})^{3 \beta-\alpha/2}} \frac {1} {(1+(\frac{r}{r_s})^3)^{\epsilon/3}}  \ , 
\end{equation}

\noindent where $n_0$, $r_c$, $\beta$, $\alpha$, $r_s$ and $\epsilon$ are fit parameters. In each iteration of the fitting process, and for a given radial bin bound by radii $r_i$ and $r_{i+1}$, we integrate $n^2_e(r)$ over the volume of the cylindrical shell defined by the above two radii and extending along the line of sight from $-3$Mpc to $+3$Mpc. The set of integrals from all radial bins is then compared to the corresponding SB values to determine the shape parameters in the right hand side of Equation \ref{eq:vikhnofr}.

This procedure provides the shape but not the normalization, $n_0$, of the density profiles. Computing $n_0$ requires knowledge of the emission integral measure, $EI = \int{ n_e n_p dV }$, where $n_p$ is the proton number density. This integral quantity, $EI$, is simply proportional to the normalization of the APEC spectral model, which we use to model ICM emission. The APEC model \citep{smith01} is fit to the spectra of multiple radial bins around the center of emission. The normalization of the spectral model of the $i^{th}$ radial bin, $K_i$, is related to $EI_i$ through

\begin{equation}
K_i = \frac{0.82 \ 10^{-14}}{4\pi D_A^2 (1+z)^2} \ n_0^2 V_i \ ,
\end{equation}

\noindent where $z$ is the cluster redshift and $D_A$ its angular diameter distance. We define $V_i$ as the spatial integral of $(n/n_0)^2$ over the cylinder defined by the $i^{th}$ radial bin and bounded along the line of sight direction by $\ell= \pm 20$Mpc. We have assumed above that $n_e = 0.82n_p$, which is suitable for typical ICM conditions. The best-fit $n_0$ is determined by minimizing

\begin{equation}
\chi^2_{n_0} = \sum_{i}{ \frac{ \left( K_i - C_D V_i n_0^2 \right)^2 }{\delta K_i^2} }
\label{delchin0}
\end{equation}

\noindent with respect to $n_0^2$, where $\delta K_i$ is the uncertainty on $K_i$ and $C_D \equiv 0.82 \ 10^{-14} / 4\pi D_A^2 (1+z)^2$.

\subsubsection{Temperature Profile, $kT(r)$}
\label{sec:ktofr}

To compute the temperature profiles, we again construct radial bins, and fit their spectra, resulting in a projected temperature profile. We then deproject the above temperature profile, using similar methods to those used to deproject the density profile. We choose the size of the radial bins in order to include the lowest number of counts necessary for a temperature determination, with ${10\%}$ uncertainty. This count number, $N_{nec}$, which is a function of temperature and background count fraction, is estimated by simulating spectra of different total counts, temperatures and background fractions, $f_{bg}$, to find the necessary counts for a ${ 10\% }$ uncertainty on temperature \citep{elkho12}. We find that $N_{nec}$ can approximately be fit by

\begin{equation}
N_{nec}(kT, f_{bg}) = 500 \times 10^{1.976f_{bg}} \times \left( \frac{kT}{2\mbox{keV}} \right) ^{1.7} \ .
\label{eq:nneckt}
\end{equation}

We extract spectra of at least $N_{nec}$ counts and fit their spectra to an absorbed APEC model in CIAO's tool, \textit{Sherpa}. We create a spectrum from each observation that partially or wholly covers the annulus corresponding to a radial bin, and simultaneously fit these spectra. We link the temperatures and metal abundances of these spectra, across the multiple observations, during the fit. Normalizations are only linked for spectra that cover more than 95\% of the solid angle of the annular region. Metallicity is left as a free parameter, while the hydrogen column density, $n_H$, is fixed to the values from the LAB dataset of \citet{kalbe05}. We also include in our online dataset the results of analysis with \citet{dicke90} $n_H$ measurements. We find that leaving $n_H$ as a free parameter returns unreasonable best-fit values on both $n_H$ and temperature. However, for Abell 478, which is reported to have varying $n_H$ by \citet{vikh05}, we allow $n_H$ to be free, with a minimum equal to the 21-cm-measured value.

The background spectrum is extracted from the \textit{Chandra} blank sky datasets and from source free regions around cluster observations. The background is modeled as particle background, plus X-ray background. The X-ray background is modeled as an absorbed 0.2keV APEC model, with $n_H=2.09 \times 10^{22}$ cm$^{-2}$ and an absorbed power-law component for the Cosmic X-ray Background, with index set to -1.4. These components are convolved with the instrument response for each CCD chip. The particle background is modeled as a series of Gaussian, exponential, and a power-law functions, to phenomenologically fit the remaining components of the blank sky datasets. It is not convolved with the instrument response. The overall background spectral model varies from epoch to epoch, and also depends on the CCD chip used.

We assume that the shape of the instrumental background component of the cluster observation is the same as the best-fit model from the corresponding blank sky dataset and compute its normalization as suggested in Maxim Markevitch's cookbook for treating the background data\footnote{http://cxc.harvard.edu/contrib/maxim/acisbg/COOKBOOK}. The overall normalization of the instrumental background is computed by scaling the background normalization according to the ratio of counts in the 9.5-12 keV energy range, in the cluster dataset relative to the blank sky dataset. In this manner we attempt to capture any possible change in the background normalization between different epochs. We note that datasets in our sample with OBSID between 7686 and 7701 are missing high-energy counts. Their instrumental background normalization is thus scaled simply by exposure time and solid angle.

Having used the blank sky data to constrain the instrumental background components, we proceed to fit the X-ray background from the in-field spectra. The latter are modeled with the same model described above plus an additional APEC component to account for residual cluster emission. The APEC model's temperature is fixed at the temperature measured outside a projected radius of 150kpc, using an initial simple fit.

The in-field spectra are obtained from annuli centered around the cluster center, and covering regions that are visually identified to contain mostly background X-ray counts. For Abell 119 and Abell 3571, the cluster emission covers most of the field of view (FOV.) We thus rely on the blank sky background data to model both instrumental and X-ray background components, for these two clusters.

We use CSTAT as our fit statistic, as it is more suitable for energy bins with low counts, where the more commonly used $\chi^2$ statistic introduces bias. After obtaining the best-fit temperature, we compute its uncertainty using \textit{Sherpa}'s \texttt{proj()} function, which varies temperature along a grid and searches for the best-fit at each temperature by varying the other thawed parameters. 

The spectral fitting described above, returns a best-fit \textit{projected} temperature for a given radial bin: Since the ICM is thought to be optically thin, the emission at one point on the sky is the sum of all emission from the line of sight behind that point. Thus, to compute the true three-dimensional temperature profile, we assume a flexible analytic form for $kT(r)$, vary its parameters repeatedly, projecting it along the line of sight in each iteration until the best match is found with the measured projected temperature radial profile. This fitting process is again run using \textit{Sherpa}. The projection is computed according to the prescription in \citet{mazzo04}, who show that to recover a single-temperature fit from a mixture of many temperature components, one should average these temperatures with a weighting proportional to $n^2VT^{\alpha}$, where $V$ is the volume of the region of emission. We choose $\alpha=-0.75$, as suggested by the range of values found by \citet{mazzo04} for spectra of different metallicities. The three-dimensional temperature profile is modeled as in \citet{vikh06}:

\begin{equation}
\label{eq:ktofr}
kT(r) = kT_0 \ \frac{(r/r_t)^{-a}}{\left[  1+ (r/r_t)^b  \right]^{c/b}} \  \frac{x+T_{min}/T_0}{x+1} \ ,
\end{equation}

\noindent where $x=(r/r_{cool})^{a_{cool}}$ and where $T_0$, $r_t$, $a$, $b$, $c$, $T_{min}$, $r_{cool}$ and $a_{cool}$ are fit parameters. The number of free parameters depends on the number of available temperature measurements.

Computing $kT(r)$ and $n_e(r)$ gives us the necessary quantities to measure $R_{500}$, the gas mass within $R_{500}$, which we call $M_{gas}$, and the total gravitational mass within the same radius, $M_{500}$. As described in \citet{elkho12}, we do so using an iterative scheme, since the 3 quantities are related. We use the $M-Y_X$ relation of \citet{kravt06} to relate $M_{500}$ to our measurables, as $Y_X \equiv kT_X M_{gas}$.

\subsubsection{Uncertainties}
\label{sec:uncert}

The uncertainties on the entropy profile of each cluster are estimated by generating a set of $S(r)$ models, which are allowed by the data and their uncertainties, as descibed here. For temperature radial profiles, using the temperature measurements in each radial bin, and their error estimate, we randomly generate new ``fake" datasets, and fit them one at a time. To generate a fake temperature measurement for each radial bin, we draw its value from a random distribution designed to capture the asymmetric uncertainties obtained on the bin's best-fit temperature. This probability distribution is a piece-wise function of 2 Gaussian distributions on either side of the best-fit temperature, with the standard deviations equal to the measured 1-$\sigma$ upper and lower uncertainties. The latter are not in general equal to each other. Once a complete radial temperature profile is generated over the entire available radial range, we fit it with the same model in Equations \ref{eq:ktofr}, and repeat this analysis for 400 iterations.

The same analysis is repeated for the density profiles, where surface brightness measurements are similarly perturbed for 300 iterations according to their uncertainties. The uncertainties in this case are assumed to be symmetric, and the fake surface brightness measurements are drawn from a Gaussian distribution.

To compute the uncertainty on the entropy profile, we compute a set of entropy profiles from pairing different temperature and density profiles, from the above generated sets. We iterate through all 400 temperature profiles. For each temperature profile, we iterate through 10 density profiles computing a temperature profile $S(r)=kT(r)/n_e(r)^{2/3}$, in each iteration. We ensure to choose different density profiles, from one temperature profile to the next, until all 300 profiles are used, at which point we start from the beginning of the density profiles list. The result is an ensemble of 4000 entropy profiles, which we use to find the distribution of entropy values at any given radius.

\subsubsection{Metallicity Profile Calculation Method}
\label{sec:zmethod}

To compute the metallicity profiles, we take a similar approach to that used to make the temperature profiles. First, using simulated spectra, we estimate the minimum necessary counts, $N_{nec}^Z( kT, Z, f_{bg} )$, to obtain a 20\% uncertainty on the best-fit metallicity. In this case, $N_{nec}^Z( kT, Z, f_{bg} )$ does not have a simple analytical form as its counterpart for temperature measurement, but is rather estimated from a weighted average of $N_{nec}^Z$ values estimated for the $kT$, $Z$ and $f_{bg}$ values that were simulated \citep{elkho12}. Then, using the derived $N_{nec}^Z$, we extract spectra in radial bins using the same bins used fo the $kT(r)$ profile calculation, and joining them whenever more counts are needed for a 20\%-uncertainty temperature estimate. We take the maximum radius of extraction to be $R_{500}$.

For spectral fitting, we again model both source emission from the ICM and background. For the background spectra, we use the same best-fit parameters found in the $kT(r)$ analysis, above. The background normalization is computed using the same method as in the $kT(r)$ analysis, described in Section \ref{sec:ktofr}. Cluster emission is modeled using a 1-temperature model, and using a 2-temperature model, where the cooler component's temperature is set to one half of the value of the hotter component's temperature.

We compute the uncertainty on the metallicity in each bin using \textit{Sherpa}'s \texttt{proj()} function. From the obtained metallicity profile, we characterize the metallicity of each cluster by two global quantities, $\bar{Z}_{mid}$ and $\bar{Z}_{in}$. We define 

\begin{equation}
\bar{Z}_{mid} = \frac {\sum_{0.15<r_i<0.3R_{500}}{ Z_i M_{gas,i} }} {\sum_{0.15<r_i<0.3R_{500}}{ M_{gas,i} }} \ ,
\end{equation}

\noindent which is the gas-mass-weighted metallicity over all shells in the range $0.15<r<0.3R_{500}$. Here, $Z_i$ and $M_{gas,i}$ are, respectively, the metallicity and the gas mass in the $i^{th}$ radial bin. In other words, $\bar{Z}_{mid}$ traces the total iron mass, $M_{Fe}^{mid}$, in the region $0.15R_{500}<r<0.3R_{500}$, according to

\begin{equation}
M_{Fe}^{mid} = A_{Fe} \bar{Z}_{mid} M_{gas}^{mid} \ ,
\end{equation} 

\noindent where $A_{Fe}=0.0019$ is the solar abundance of iron by mass, according to the photospheric measurements in \citet{ander89} assumed for our spectral analysis, and $M_{gas}^{mid}$ is the gas mass contained in the same region. Similarly, 

\begin{equation}
\label{eq:zavgin}
\bar{Z}_{in} = \frac {\sum_{r_i<0.15R_{500}}{ Z_i M_{gas,i} }} {\sum_{r_i<0.15R_{500}}{ M_{gas,i} }} \ .
\end{equation}

Hereafter, $\bar{Z}_{mid}$ will be used as a measure of the metallicity of the bulk of a cluster, while $\bar{Z}_{in}$ will be used as a measure of the core metallicity in clusters.

\section{Results}

We make our data available on the FTP site\footnote{The file named ``README" within this FTP site details the content of the data.}

\begin{center}
\texttt{ftp://space.mit.edu/pub/tamer/ebc2015/} \ .
\end{center}

Appendix \ref{app:sofr} contains the plot of the entropy profile for each cluster, while the individual metallicity radial profiles are shown in Appendix \ref{app:zofr}. 

We first note that 4 clusters of our sample deviate considerably from spherical symmetry simply based on their surface brightness image. These clusters, Abell 754, Abell 2256, Abell 3376 and Abell 3667, are known to be undergoing merging events. We exclude the disturbed clusters from our analysis, except when noted.

Table \ref{tab:masses} in Appendix \ref{app:masses} shows the measured values of $kT_X$, $R_{500}$, $M_{500}$ and $M_{gas}$ for each cluster.

\subsection{Entropy Profiles}

Our best-fit results for density and temperature radial profiles are shown in Tables \ref{tab:nofr} and \ref{tab:ktofr}, respectively, in Appendix \ref{app:profiles}.

\begin{figure*}[ht]
\centering
\hspace{-3em}
% plot generated by ~/caviar/research/acchif3/sz6_sofr2.pro
%\makebox[\columnwidth]{}
\scalebox{0.8}{\includegraphics*{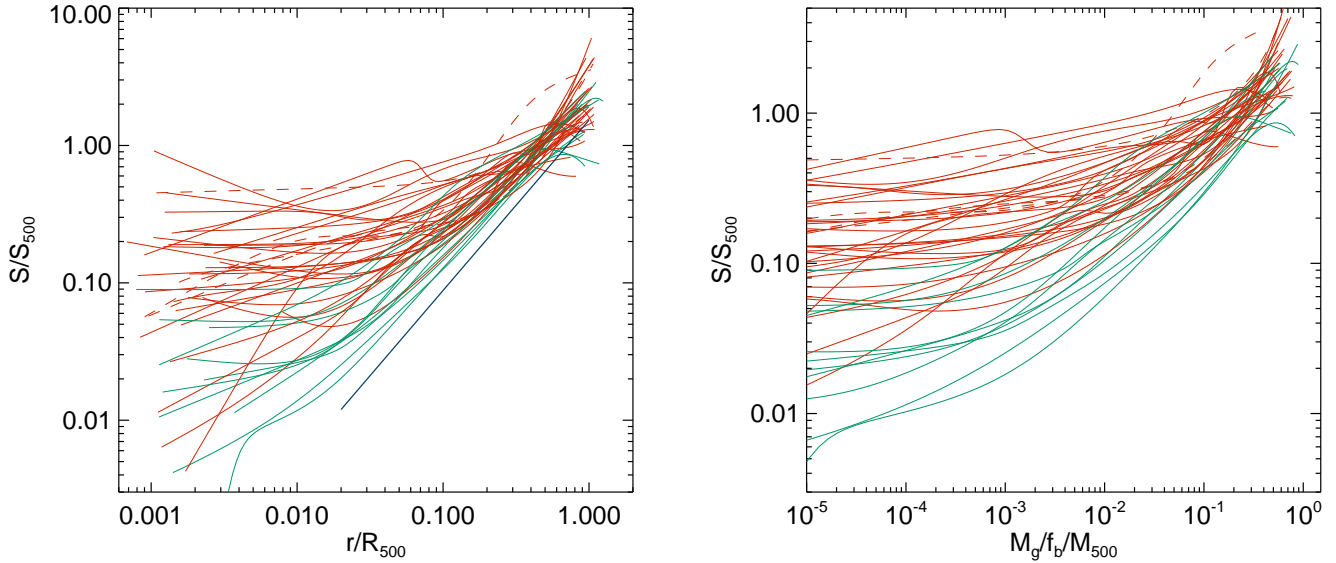}}
\caption{Plot of all computed entropy profiles. \textit{Left:} scaled entropy, $S/S_{500}$, as a function of scaled radius, $r/R_{500}$. \textit{Right:} scaled entropy as a function of enclosed gas mass fraction, $F_g$. Turquoise curves are for CC clusters, while red curves are for NCC clusters. The dark blue curve is the power law describing the Voit 2005 entropy profile found in hydrodynamical simulations. Dashed curves are for the 4 morphologically disturbed clusters Abell 754, Abell 2256, Abell 3376 and Abell 3667.}
\label{fig:sofrall}
\end{figure*}

We overplot the computed entropy profiles for all of our sample's clusters in Figure \ref{fig:sofrall}. On the left panel, we plot entropy as a function of radius, which we normalize with respect to $R_{500}$. On the right panel, we plot entropy as a function of enclosed gas mass fraction, $F_g \equiv M_g/(f_bM_{500})$, where $M_g$ is the interior gas mass, and $f_b$ the universal baryon fraction with respect to all matter, i.e. $f_b=\Omega_b/\Omega_m$. We use $F_g$ to plot entropy profiles because this is the variable used in a Lagrangian description of the entropy distribution in clusters \citep[see e.g.][]{tozzi01,voit03,nath11}. The entropy in both panels is normalized with respect to $S_{500}$ \citep{voit03}, the characteristic entropy of the cluster at $R_{500}$:

\begin{equation}
\label{eq:s500}
S_{500} \equiv \frac{GM_{500}\mu m_p}{2 R_{500} \left[ 500 f_b \rho_c(z)/\mu_e m_p \right]^{2/3}} \ ,
\end{equation}

\noindent where $\mu$ and $\mu_e$ are the mean number of nucleons per particle and per electron, respectively, $m_p$ is the proton mass and $\rho_c(z)$ is the Universe's critical density at the redshift of observation, $z$. The characteristic entropy, $S_{500}$ at an overdensity $\delta=500$ is simply the entropy obtained using the characteristic temperature at $\delta=500$, which is the equivalent of the virial temperature but defined for $R_{500}$ instead of the virial radius,

\begin{equation}
kT_{500} \equiv \frac{GM_{500}\mu m_p}{2R_{500}} \ ,
\end{equation}

\noindent and using the average electron density inside $R_{500}$,

\begin{equation}
\bar{n}_e = 500 f_b \rho_c(z)/\mu_e m_p \ .
\end{equation}

\noindent It represents the entropy scale set by gravity in the self-similar picture.

The dark blue line in Figure \ref{fig:sofrall} represents a model of the entropy profile of a cluster generated from gravitational collapse alone, which was calculated with hydrodynamical AMR simulations in \citet{voit05}. Voit's entropy profile is approximated analytically as a power law,

\begin{equation}
\label{eq:voit05}
S_V(r) = 1.53 S_{500} \left( \frac{r}{R_{500}} \right)^{1.24} \ ,
\end{equation}

\noindent and is valid for radii larger than approximately $0.2R_{500}$. We employ the conversion used in \citet{point08} to express Equation \ref{eq:voit05} in terms of $S_{500}$ and $R_{500}$, as opposed to the measurements at an overdensity of 200, presented in \citet{voit05}.

Turquoise curves in Figure \ref{fig:sofrall} represent cool core clusters (CC), while red curves are for non-cool core clusters (NCC). We use the surface brightness concentration, $c_{SB}$, introduced in \citet{santo08} to quantify the cool core state of a cluster. The parameter $c_{SB}$ is defined as the ratio of the surface brightness within 40 kpc to that within 400kpc of the \textit{peak} of the emission\footnote{For calculating $c_{SB}$, we use the emission peak as the cluster center, unlike in previous analysis where the emission is used instead.}. The values of $c_{SB}$ are shown in Table \ref{tab:morph}. We define CC clusters as clusters with $c_{SB}>11$, while NCC cluster have $c_{SB}<11$.

The first observation to make is that for most of the studied radial range, all entropy profiles lie above Voit's 2005 gravitationally induced entropy model. This result has been known in the literature \citep[e.g.][]{ponma99,pratt10}, and the additional entropy in the observations has been attributed to non-gravitational processes, such as winds and AGN heating. Second, some of the entropy profiles in Figure \ref{fig:sofrall} show a decrease or a flattening starting at a radius between 0.3 and $0.8R_{500}$. This is due to a decrease in measured temperature towards the outskirts of many clusters, which is not matched by a steep enough decrease of measured density with radius. Such a configuration of large amounts of lower-entropy gas at larger radii is not physically stable. There are two sources of systematic error that can be contributing here to give erroneous temperature and density measurements. First, systematics in our estimate of the level of X-ray background will translate into an error in the estimate of the cluster surface brightness at these large radii, introducing a bias to the inferred density, and also biasing outer temperature estimates. Second, the deprojection method of \citet{mazzo04}, which we employ to deproject the measured 2D temperature profile to a 3D $kT(r)$, is known to be less accurate when there is significant contribution to the emission from spectral components with temperatures smaller than $\sim 3$keV \citep{mazzo04,vikh06proj}.

In addition, we show in Figure \ref{fig:sofr}, in Appendix \ref{app:sofr}, the individual entropy profiles we compute. The estimate of the 1-sigma range of entropy at each radius is represented by the turquoise shaded region. The uncertainty in the measured temperatures, which translates into an uncertainty in the parameters of the temperature radial profile, is the main contributor to the uncertainty in the entropy profile. By comparison, the contribution of the density uncertainty to the entropy uncertainty is much smaller.

The light gray error bars, in Figure \ref{fig:sofr}, represent the entropy profiles measured in the ACCEPT study by \citet{cavag09}. Our entropy profiles agree, in general, with the ACCEPT entropy profiles, where they overlap. However, we extend our entropy profiles to larger radii, where we model both density and temperature.

\subsection{Metallicity Profiles and Global Measurements}

\subsubsection{Profiles}

\begin{figure}[h]
\centering
% plot generated by ~/caviar/research/acchif3/plotzofr_all5.pro
\scalebox{0.7}{\includegraphics{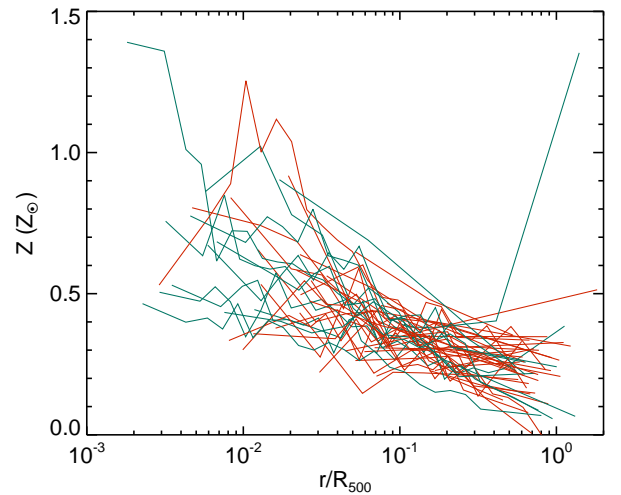}} %*[0.15in,2.5in][8in,10in]
\caption{Superimposed metallicity profiles of clusters. Red curves are for NCC clusters, while turquoise curves are for CC clusters.}
\label{fig:zofrall}
\end{figure}

We overplot all obtained metallicity profiles in Figure \ref{fig:zofrall}. The dispersion in the values of observed cluster metallicities decreases for radii larger than $0.1-0.2R_{500}$, despite the larger uncertainty associated with measured metallicities at these high radii. One metallicity measurement seems to be exceptionally larger, at large radius, as seen in Figure \ref{fig:zofrall}. This is the last metallicity measurement for Abell 2204, where modeling the background is likely suffering from systematics, despite an acceptable fit statistic. The outermost spectra for Abell 2204 are found to have excess low-energy counts, which are mostly fit by our galactic X-ray background component.

We also present in Appendix \ref{app:zofr} plots of our measured metallicity profiles, individually, with their estimated uncertainty. The two vertical lines in Figure \ref{fig:zofr} are drawn at radii $0.15R_{500}$ and $R_{500}$, for those clusters where we can measure $M_{500}$ and $R_{500}$ using our iterative method.

\subsubsection{Core Metallicity, $\bar{Z}_{in}$}
\label{sec:zin}

\begin{figure*}[ht]
\centering
% plot generated by ~/caviar/research/acchif3/z_m500_ktx_mid3.pro
\scalebox{0.8}{\includegraphics{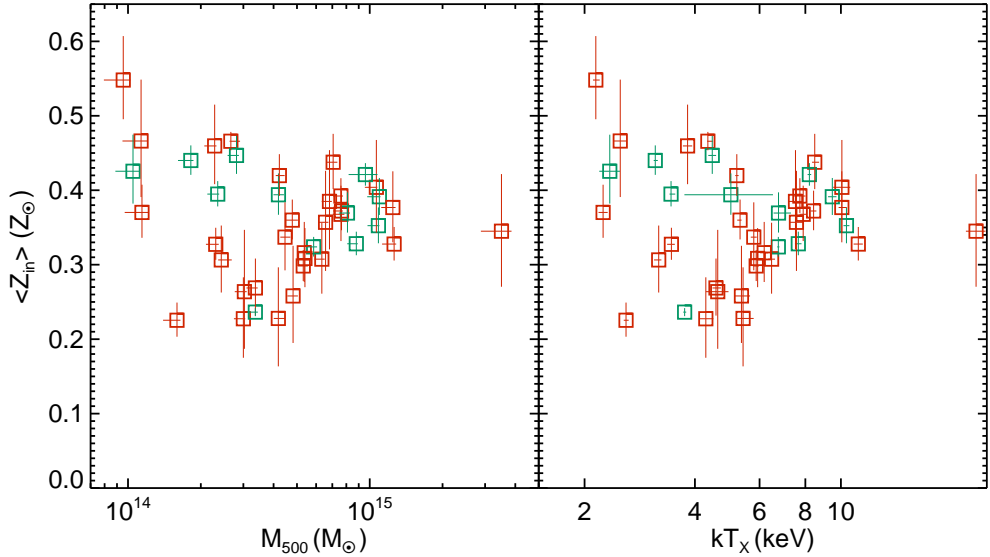}}
\caption{Core metallicity as a function of cluster mass or temperature. \textit{Left:} $\bar{Z}_{in}$ vs. $M_{500}$. \textit{Right:} $\bar{Z}_{in}$ vs. $kT_X$. Red points are for NCC clusters, while turquoise points are for CC clusters.}
\label{fig:zin_m500}
\end{figure*}

The measured values of $\bar{Z}_{in}$ are shown in Table \ref{tab:globalz}, in Appendix \ref{app:globalz}. We plot $\bar{Z}_{in}$ as a function of cluster mass, and as a function of the global temperature measurement, $kT_X$, in Figure \ref{fig:zin_m500}. Figure \ref{fig:zin_m500} suggests that low-mass clusters ($M_{500}<3.5 \times 10^{14} M_\odot$) exhibit more dispersion in $\bar{Z}_{in}$ than high-mass clusters do. 

To quantify this latter observation, we show in Table \ref{tab:zinstats} the error-weighted mean, and the standard deviation of the values of $\bar{Z}_{in}$, for clusters grouped by mass. The results for the 1-temperature fit are labeled ``1T", and those for the 2-temperature fit ``2T." The standard deviation of $\bar{Z}_{in}$ for clusters with $M_{500}>3.5 \times 10^{14} M_\odot$ is 2.4 to 2.9 times smaller than that for clusters with $M_{500}<3.5 \times 10^{14} M_\odot$, for the 1T and the 2T models, respectively. The value of $\chi^2$ for each subset shows whether the dispersion for low-mass clusters is solely the result of measurement uncertainties. Here,

\begin{equation}
\chi^2 = \sum_i \frac{ \left( \bar{Z}_{in,i} - \langle \bar{Z}_{in} \rangle \right)^2 }{\delta \bar{Z}_{in,i}^2} \ ,
\end{equation}

\noindent where the sum is over the sample of clusters denoted by $i$, $\bar{Z}_{in,i}$ is the $i^{th}$ cluster's central metallicity, $\delta \bar{Z}_{in,i}$ is its uncertainty and $\langle \bar{Z}_{in} \rangle$ the sample mean. 

For low-mass clusters, we find that $\chi^2$ is more than 30 times the number of degrees of freedom, for both 1T and 2T fits. This confirms that the dispersion seen in low-mass clusters --- in the range $(0.12-0.15)Z_\odot$ --- is not driven by measurement uncertainties. On the contrary, the dispersion of $\bar{Z}_{in}$ in high-mass clusters has a more significant contribution from measurement uncertainties, despite being much smaller at $\sim 0.05Z_\odot$.

\begin{table}[here]
\begin{center}
% values taken from output of calc_zin_chi2b_postref.pro
\caption{Statistics for $\bar{Z}_{in}$.\label{tab:zinstats}}
\begin{tabular}{ @{}lcccc@{} }
\toprule
                                                                   & Model    &   Mean &     Std. Dev. & $\chi^2$/dof         \\ 
 & & ($Z_\odot$) & ($Z_\odot$) & \\ \midrule

\multicolumn{1}{l}{\multirow{2}{*}{$M_{500}<3.5 \times 10^{14} M_\odot$}}  & 1T & 0.30    &      0.12     &  33. (490./15)       \\
\multicolumn{1}{l}{}                                                     &  2T & 0.25    &      0.15     &  31. (470./15)        \\ \midrule

\multicolumn{1}{l}{\multirow{2}{*}{$M_{500}>3.5 \times 10^{14} M_\odot$}}  & 1T & 0.36    &     0.050     &   2.6 (64./25)       \\
\multicolumn{1}{l}{}                                                     & 2T  & 0.35    &     0.052     &   1.2 (29./25)        \\ \bottomrule

\end{tabular}
\tablecomments{Error-weighted mean, standard deviation of $\bar{Z}_{in}$ and $\chi^2/dof$ with respect to calculated mean. Results shown for clusters with $M_{500}$ smaller than and larger than $3.5 \times 10^{14} M_\odot$, and for fits using the 1T and the 2T spectral models}
\end{center}
\end{table}

The above results are unchanged when we include the $\bar{Z}_{in}$ measurements from the high-mass, asymmetrical clusters Abell 754, Abell 2256 and Abell 3667, which have $\bar{Z}_{in}=0.369^{+0.036}_{-0.035}$, $0.389^{+0.065}_{-0.061}$ and $0.345^{+0.012}_{-0.011} Z_\odot$, respectively. These clusters were excluded from the above analysis based on their asymmetric morphologies. The 1T dispersion of $\bar{Z}_{in}$, over the high-mass subset decreases to $0.047 Z_\odot$, when the above 3 clusters are included, while the 2T dispersion does not change significantly.

\begin{figure}[h]
\centering
% plot generated by ~/caviar/research/acchif3/plotzofr_all4_bymass2.pro
\scalebox{0.7}{\includegraphics{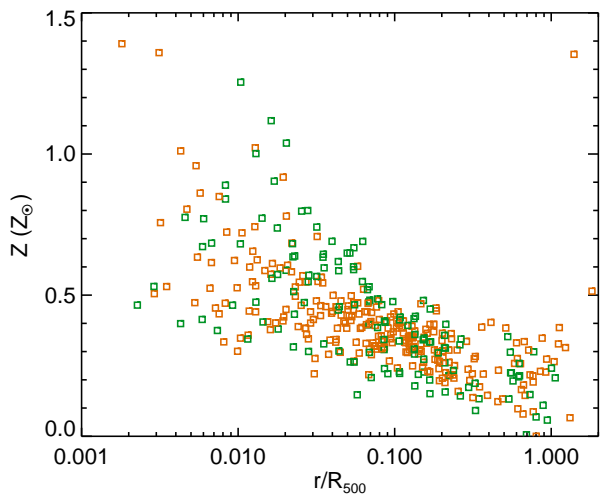}} %*[0.15in,2.5in][8in,10in]
\caption{Superimposed metallicity profiles of clusters. Orange data points are for clusters with mass $M_{500}>3.5 \times 10^{14} M_\odot$, while green data points correspond to clusters with $M_{500}<3.5 \times 10^{14} M_\odot$}
\label{fig:zofrall_bymass}
\end{figure}

To emphasize the difference between low- and high-mass clusters metallicity dispersions, we show in Figure \ref{fig:zofrall_bymass} the superimposed metallicity radial profiles of all clusters in our sample, where we differentiate between the two subsets by color. Green data points represent low-mass clusters, while orange data points represent larger clusters. We can see in this plot that large clusters' metallicity profiles are less scattered than low-mass clusters'. This translates in the different observed dispersions of $\bar{Z}_{in}$ seen in Figure \ref{fig:zin_m500}.

% CC vs NCC
Comparing CC and NCC clusters, we find that for CC clusters $\bar{Z}_{in}=0.30 \pm 0.10 Z_\odot$, while for NCC clusters $\bar{Z}_{in}=0.37 \pm 0.080 Z_\odot$. We find that $\bar{Z}_{in}$ is larger for NCC clusters compared to CC clusters in this sample, despite the difference being within the measured dispersions of both quantities. This contrasts to the metal excess measured in the centers of CC cluster in e.g. \citet{degra01}.

\subsubsection{Outer Metallicity, $\bar{Z}_{mid}$}

The measured values of $\bar{Z}_{mid}$ are shown in Table \ref{tab:globalz}, in Appendix \ref{app:globalz}. In Figure \ref{fig:zmid_m500}, we show a plot of $\bar{Z}_{mid}$, which measures the average metallicity outside the core, as a function of the total mass, $M_{500}$, and as a function of $kT_X$. In Table \ref{tab:zmidstats}, we show the statistics for the distribution of $\bar{Z}_{mid}$ values. In the case of outer metallicity, we no longer detect a clear difference in the dispersions of high- and low-mass clusters. However, when we compare $\bar{Z}_{mid}$ to $\bar{Z}_{in}$, we find that for each of the low- and high-mass cluster samples, $\bar{Z}_{mid}$ values are smaller than $\bar{Z}_{in}$ values. For high-mass clusters, for example, the mean and standard deviation for $\bar{Z}_{mid}$ are $(0.27 \pm 0.073)Z_\odot$, while for $\bar{Z}_{in}$ they are $(0.36 \pm 0.050)Z_\odot$. This points to a decrease in the iron mass fraction as we move from the core region, $r<0.15R_{500}$, to the outer region $0.15<r<0.3R_{500}$. This decrease is however within the measured dispersions of $\bar{Z}_{mid}$ and $\bar{Z}_{in}$ and is also found for low-mass clusters.

\begin{figure}[ht]
%\centering
\hspace{-2em}
% plot generated by ~/caviar/research/acchif3/z_m500_ktx_mid3.pro
\scalebox{0.54}{\includegraphics{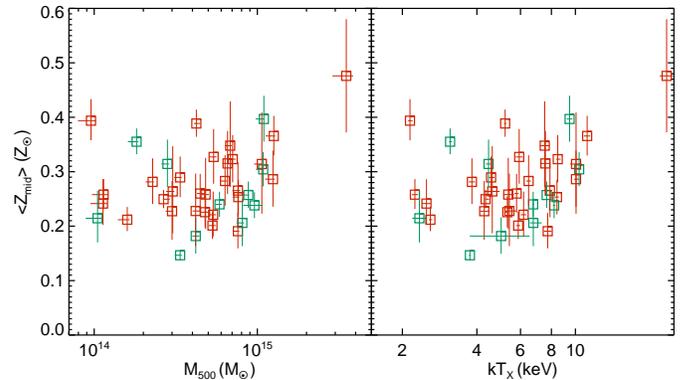}}
\caption{Outer metallicity as a function of cluster mass or temperature. \textit{Left:} $\bar{Z}_{mid}$ vs. $M_{500}$. \textit{Right:} $\bar{Z}_{mid}$ vs. $kT_X$. Red points are for NCC clusters, while turquoise points are for CC clusters.}
\label{fig:zmid_m500}
\end{figure}

\begin{table}[here]
\begin{center}
% values taken from output of calc_zmid_chi2b_postref.pro
\caption{Statistics for $\bar{Z}_{mid}$.\label{tab:zmidstats}}
\begin{tabular}{ @{}lcccc@{} }
\toprule
                                                                   & Model    &   Mean &     Std. Dev. & $\chi^2$/dof         \\ 
 & & ($Z_\odot$) & ($Z_\odot$) & \\ \midrule

\multicolumn{1}{l}{\multirow{2}{*}{$M_{500}<3.5 \times 10^{14} M_\odot$}}  & 1T & 0.22    &      0.079     &  10. (150./15)       \\
\multicolumn{1}{l}{}                                                     &  2T & 0.20    &      0.085     &  9.1 (140./15)        \\ \midrule

\multicolumn{1}{l}{\multirow{2}{*}{$M_{500}>3.5 \times 10^{14} M_\odot$}}  & 1T & 0.27    &     0.073     &   3.2 (79./25)       \\
\multicolumn{1}{l}{}                                                     & 2T  & 0.27    &     0.074     &   2.3 (56./25)        \\ \bottomrule

\end{tabular}
\tablecomments{Error-weighted mean, standard deviation of $\bar{Z}_{mid}$ and $\chi^2/dof$ with respect to calculated mean. Results shown for clusters with $M_{500}$ smaller than and larger than $3.5 \times 10^{14} M_\odot$, and for fits using the 1T and the 2T spectral models}
\end{center}
\end{table}

\subsection{Metallicity-Entropy Relation}
\label{sec:zxs}

One direct approach to look for a relation between pre-enrichment and pre-heating is to look for a correlation between the ICM non-gravitational entropy and the ICM bulk metallicity measured outside the central region of the cluster, $\bar{Z}_{mid}$. In addition, we also consider the relation between non-gravitational entropy and $\bar{Z}_{in}$. We use the ratio of measured entropy to the expected gravitational entropy, $S_{grav}$, to probe the amount non-gravitational entropy. We define the scaled entropy $x_s \equiv S/S_{grav}$, which we use as a measure for any non-gravitational entropy, and discuss our assumptions on $S_{grav}$, below.

The metallicity-entropy relation is studied with entropy measured at several locations in the clusters. First, we consider entropy measurements at fixed $R_{500}$-scaled radii. This is justified because the gravitational entropy model of \citet{voit05} scales self-similarly, and is given in terms of a profile which is a function of $r/R_{500}$. In this case, we define $S_{grav}$ as the expected gravitational entropy from Voit's model, $S_{grav}(r) = S_V(r)$. Second, we take a Lagrangian approach and study entropy at a fixed interior gas mass fraction, $F_g \equiv M_g/(f_bM_{500})$, where $M_g$ is the interior gas mass \citep[see e.g.][]{tozzi01,voit03,nath11}. This can be useful because buoyancy tends to order the ICM such that low-entropy gas finds its way to the bottom of the cluster potential, while high-entropy gas rises to large radii. In this latter case, we simply use $S_{grav}=S_{500}$ to scale the entropy, to avoid using a specific model of entropy dependence on $F_g$, while still capturing the $S_{500}$ scaling expected in self-similar galaxy clusters.

In the analysis below not all clusters are included for each measurement. The first filter we apply is to exclude 4 clusters, which are visually judged to greatly deviate from spherical symmetry. These are Abell 754, Abell 2256, Abell 3376 and Abell 3667. In addition, in the successive measurements at different radii, below, we only include a cluster at a certain radius if the size of the FOV is larger than the radius of interest.

\begin{figure*}[ht]
\centering
\hspace{-2em}
% plot generated by ~/caviar/research/acchif3/sz6midmult2lin4.pro
%\scalebox{0.7}{\includegraphics*[0.2in,0.1in][7in,3.5in]{sz6in_r1.0.eps}}
\scalebox{0.6}{\includegraphics*{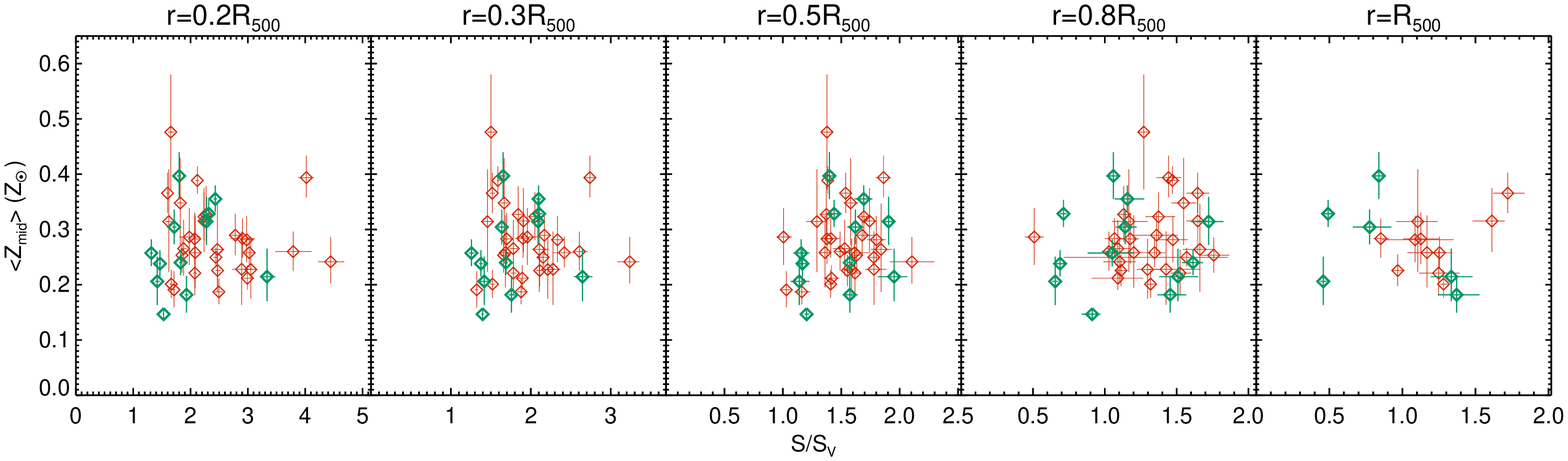}}
\caption{Plot of gas mass-weighted metallicity $\bar{Z}_{mid}$, vs. the scaled entropy, $S/S_V$. Panels from left to right represent the cases where entropy is measured at $r=0.2R_{500}$, $r=0.3R_{500}$, $r=0.5R_{500}$, $r=0.8R_{500}$ and $r=R_{500}$. Only clusters with \textit{Chandra} coverage at each of these radii are represented in the corresponding panel. Turquoise data points are for CC cluster, while red points represent NCC clusters. Abell 400 is the outlying cluster with large metallicity at $\bar{Z}_{mid} \sim 0.6 Z_\odot$.}
\label{fig:szmidr}
\end{figure*}

\begin{figure*}[ht]
\centering
\hspace{-2em}
% plot generated by ~/caviar/research/acchif3/sz6midmult2lin4.pro
%\scalebox{0.7}{\includegraphics*[0.2in,0.1in][7in,3.5in]{sz6in_r1.0.eps}}
\scalebox{0.6}{\includegraphics*{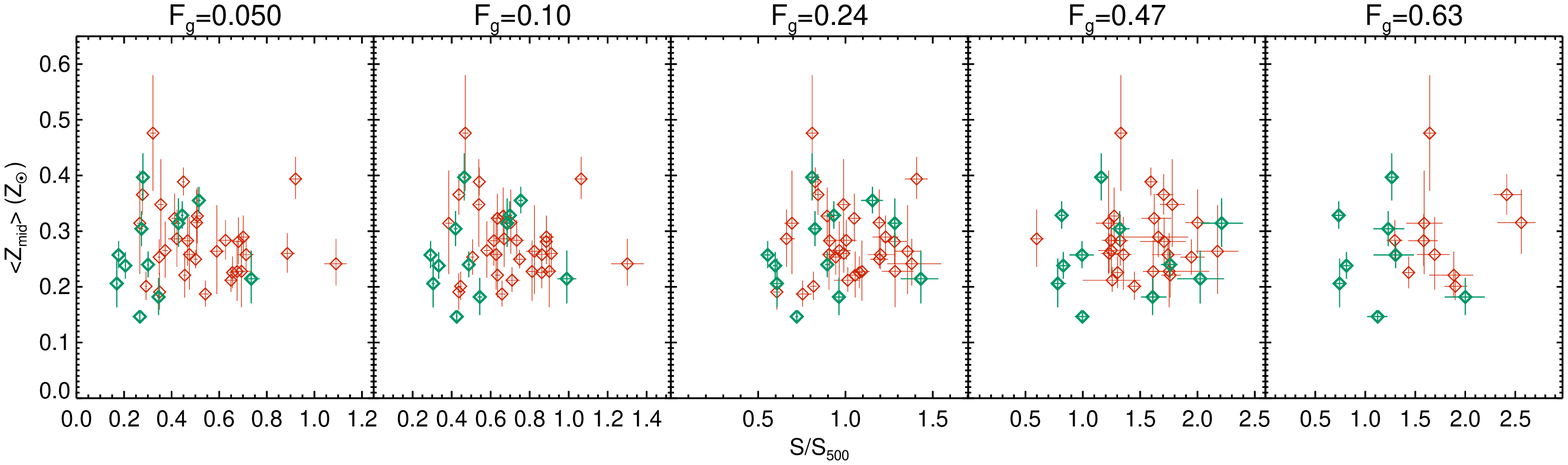}}
\caption{Plot of gas mass-weighted metallicity $\bar{Z}_{mid}$, vs. the scaled entropy, $S/S_V$. Panels from left to right represent the cases where entropy is measured at $F_g=0.050$, $F_g=0.10$, $F_g=0.24$, $F_g=0.47$ and $F_g=0.63$. Only clusters with \textit{Chandra} coverage at each of these radii are represented in the corresponding panel. Turquoise data points are for CC cluster, while red points represent NCC clusters. Abell 400 is the outlying cluster with large metallicity at $\bar{Z}_{mid} \sim 0.6 Z_\odot$.}
\label{fig:szmidfg}
\end{figure*}

We start by looking at the $\bar{Z}_{mid}-x_s$ relation at constant scaled radius. As described above, in this case, $x_s = x_s(r) = S(r)/S_{grav}(r) = S(r)/S_V(r)$. We measure entropy at $r=0.2R_{500}$, $0.3R_{500}$, $0.5R_{500}$, $0.8R_{500}$ and $1R_{500}$. Our measurements are shown in Figure \ref{fig:szmidr}. Similarly, we consider the same relation at fixed $F_g$. We choose values of $F_g$ corresponding to the sample average across all clusters, at $r=0.2$, 0.3, 0.5, 0.8 and $1R_{500}$. Table \ref{tab:rfg} shows the correspondence between scaled radius and the sample average gas mass fraction. Our metallicity-entropy measurements at constant $F_g$ are shown in Figure \ref{fig:szmidfg}.

\begin{table}[here]
\begin{center}
\caption{Cluster sample average of $F_g$ at various radii.\label{tab:rfg}}
\begin{tabular}{ @{}llllll@{} }
\toprule
$r/R_{500}$  & 0.2    &   0.3    &   0.5   &   0.8   & 1.0 \\ \midrule
$F_g$        & 0.050  &  0.10   &   0.24  &   0.47  & 0.63 \\
\bottomrule
\end{tabular}
\end{center}
\end{table}

As can be seen in Figures \ref{fig:szmidr} and \ref{fig:szmidfg}, there is no visible correlation between our estimate of non-gravitational entropy, and the bulk metal content of a cluster, as estimated by $\bar{Z}_{mid}$.

\begin{figure*}[ht]
\centering
\hspace{-2em}
% plot generated by ~/caviar/research/acchif3/sz6inmult2lin4.pro
\scalebox{0.6}{\includegraphics*{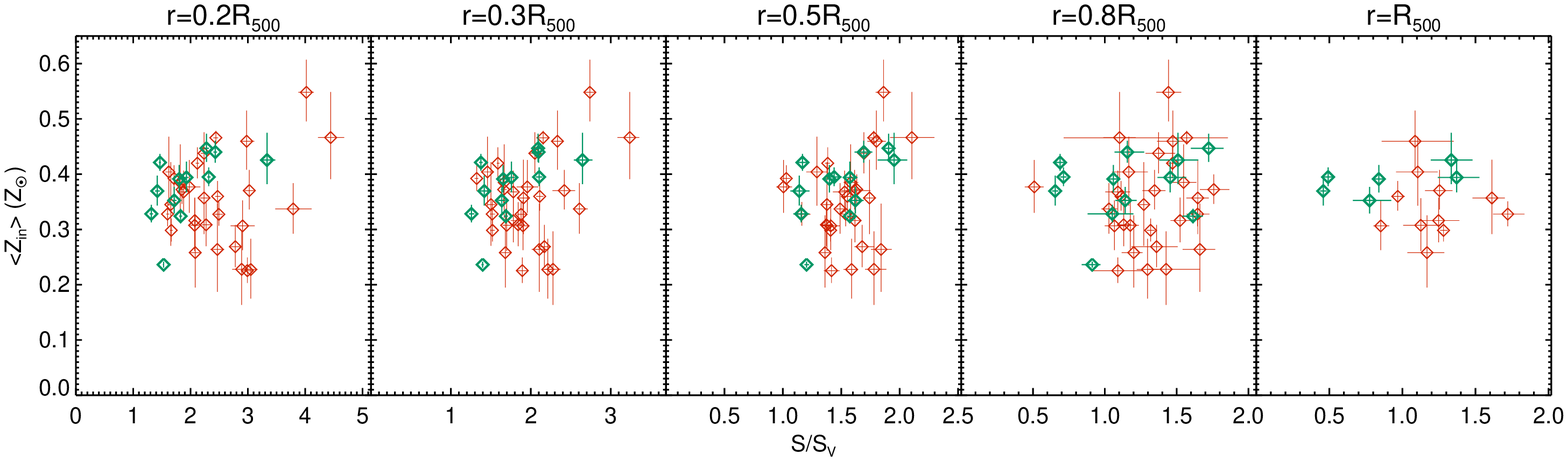}}
\caption{Plot of gas mass-weighted metallicity $\bar{Z}_{in}$, vs. the scaled entropy, $S/S_V$. Panels from left to right represent the cases where entropy is measured at $r=0.2R_{500}$, $r=0.3R_{500}$, $r=0.5R_{500}$, $r=0.8R_{500}$ and $r=R_{500}$. Only clusters with \textit{Chandra} coverage at each of these radii are represented in the corresponding panel. Turquoise data points are for CC cluster, while red points represent NCC clusters.}
\label{fig:szinr}
\end{figure*}

As for metallicity measured in the core, $\bar{Z}_{in}$, we expect low-radius metallicity measurements to probe processes that occur after the collapse of the cluster. We repeat the analysis performed above, with $\bar{Z}_{in}$ instead of $\bar{Z}_{mid}$. Figure \ref{fig:szinr} shows plots of the measured inner metallicity, $\bar{Z}_{in}$, against the ratio of measured entropy to $S_V$, at the above-mentioned scaled radii. Again, there is only weak indication of a correlation between inner metallicity and $x_s$, at smaller radii. We perform statistical analysis using a bootstrap resampling method to calculate the significance of the correlation between the various metallicity and entropy measures. The lowest obtained p-values are of 1.3\% and 2.0\% for the CC-only samples at $r=0.3$ and $0.2R_{500}$, respectively.

\begin{figure*}[ht]
\centering
\hspace{-2em}
% plot generated by ~/caviar/research/acchif3/sz6inmult2lin4.pro
\scalebox{0.6}{\includegraphics*{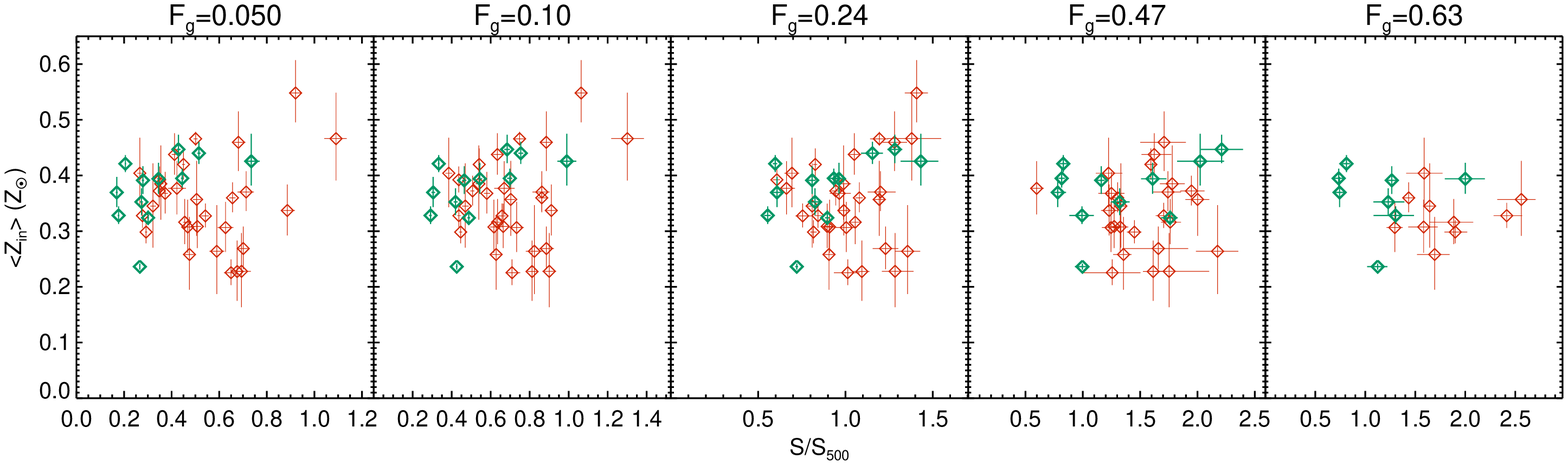}}
\caption{Plot of gas mass-weighted metallicity $\bar{Z}_{in}$, vs. the scaled entropy, $S/S_V$. Panels from left to right represent the cases where entropy is measured at $F_g=0.050$, $F_g=0.10$, $F_g=0.24$, $F_g=0.47$ and $F_g=0.63$. Only clusters with \textit{Chandra} coverage at each of these radii are represented in the corresponding panel. Turquoise data points are for CC cluster, while red points represent NCC clusters.}
\label{fig:szinmgas}
\end{figure*}

\section{Discussion}
\label{sec:disc}

\subsection{Inner Metallicity Scatter Systematics}
\label{sec:syst}

As was presented in Section \ref{sec:zin}, the core iron mass fraction, $\bar{Z}_{in}$, over our cluster sample has a different distribution for low-mass clusters than for high-mass clusters. Measurements of $\bar{Z}_{in}$ in large clusters ($M>3.5 \times 10^{14} M_\odot$) are narrowly distributed around their mean of $0.36 Z_\odot$, with a standard deviation of only $\sigma_Z = 0.050Z_\odot$. On the other hand, $\bar{Z}_{in}$ for low-mass clusters has a standard deviation of $\sigma_Z = 0.12 Z_\odot$, around a slightly lower mean value for the sample. See Table \ref{tab:zinstats} for details. The uncertainties on the individual $\bar{Z}_{in}$ measurements are too small to explain the dispersion in low-mass clusters, since $\chi^2 / \mbox{d.o.f} = 33$. This means that the observed scatter is intrinsic to the data, and not a result of measurement uncertainties. On the other hand, we calculate $\chi^2/\mbox{d.o.f}=2.6$ for $\bar{Z}_{in}$ in large clusters, indicating that measurement uncertainties contribute relatively more to the scatter, which nonetheless has a much lower value of only $\sigma_Z = 0.050Z_\odot$. 

We attempt here to understand the difference between the distribution of low- and high-mass clusters' inner metallicity values. First we check whether the observed effect is due to systematics, and then we present several physical explanations of the measurements, in the following sections. 

\begin{figure}[ht]
\centering
% plot generated by ~/caviar/research/acchif3/sz6inmult2lin3nozbin.pro
\scalebox{0.6}{\includegraphics{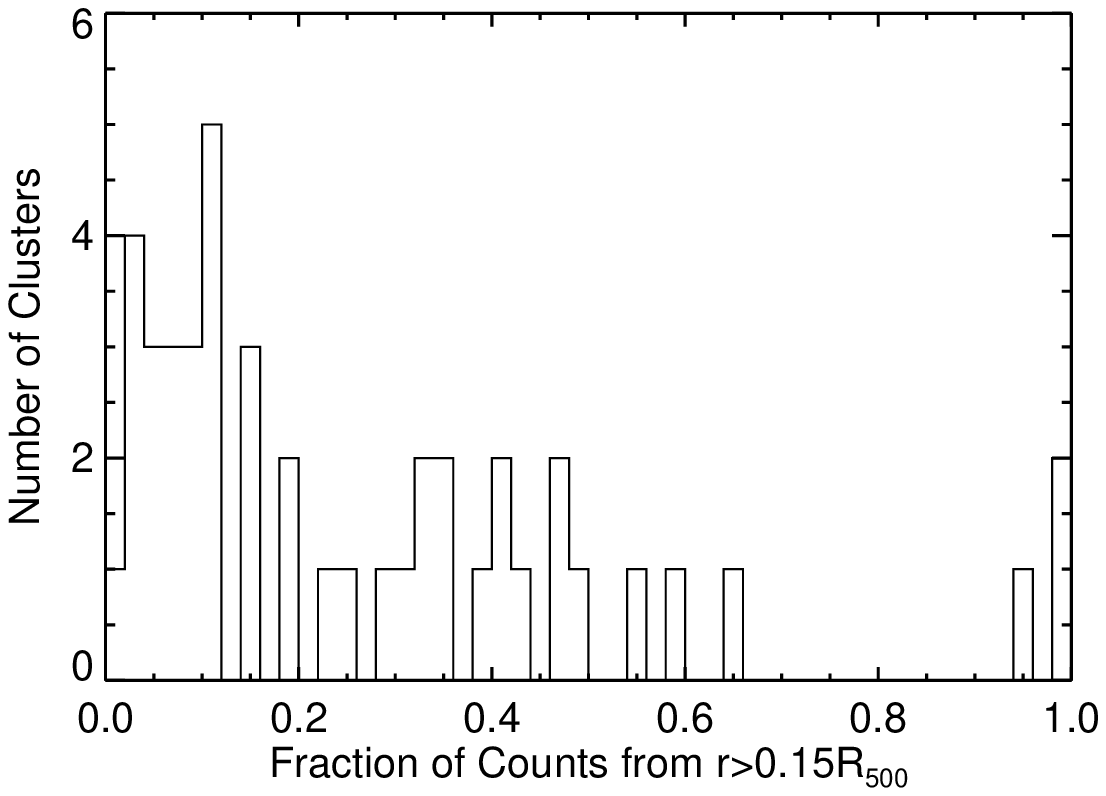}}
\caption{Distribution of the values of $f_{out}$ from the ensemble of $\bar{Z}_{in}$ measurements. See the text in Section \ref{sec:syst} for a description of $f_{out}$.}
\label{fig:fouthist}
\end{figure}

The first systematic effect to be tested concerns the inclusion of X-ray photons from the region $r>0.15R_{500}$ in the computation of $\bar{Z}_{in}$. This occurs because spectral regions for metallicity measurement are defined before $R_{500}$ is computed, while we desire for $\bar{Z}_{in}$ to measure the metallicity within 0.15$R_{500}$. Once $R_{500}$ is calculated, to compute $\bar{Z}_{in}$, we include all radial bins which overlap the disc $r<0.15R_{500}$. The last such bin will, in general, extend beyond $r=0.15R_{500}$. We address this by estimating the fraction of counts originating from the region $r>0.15R_{500}$, which are used to compute $\bar{Z}_{in}$. We denote this count fraction by $f_{out}$, and show a histogram of its distribution in our sample in Figure \ref{fig:fouthist}.

Since metallicity profiles generally decrease with radius, including emission from large radii for a given cluster might bias its $\bar{Z}_{in}$ measurement, compared to the rest of the sample. We would like to test the magnitude of this effect. We thus repeat the measurement of $\bar{Z}_{in}$ dispersion with a sample which excludes clusters with significant contribution from $r>0.15R_{500}$. We choose the cutoff value to be $f_{out}=0.25$ to capture the peak of the distribution of clusters with $f_{out}<0.25$ seen in Figure \ref{fig:fouthist}. We calculate the mean, dispersion and $\chi^2$/d.o.f statistics as was done with the complete sample. The results are displayed in Table \ref{tab:zinstats_nozbin}. We find that the difference between the high- and low-mass cluster samples still remains for dispersion and $\chi^2$/d.o.f. We thus conclude that the spectral bins' sizes do not have a large effect on this discrepancy in measured metallicity dispersions.

\begin{table}[here]
\begin{center}
\caption{Statistics for $\bar{Z}_{in}$ excluding clusters with $f_{out}>0.25$.\label{tab:zinstats_nozbin}}
% values taken from output of calc_zin_chi2b_nozbin_postref.pro
\begin{tabular}{ @{}lcccc@{} }
\toprule
                                                                   & Model    &   Mean &     Std. Dev. & $\chi^2$/dof         \\ 
 & & ($Z_\odot$) & ($Z_\odot$) & \\ \midrule

\multicolumn{1}{l}{\multirow{2}{*}{$M_{500}<3.5 \times 10^{14} M_\odot$}}  & 1T & 0.30    &      0.15     &  59. (470./8)       \\
\multicolumn{1}{l}{}                                                     &  2T & 0.26    &      0.18     &  57. (460./8)        \\ \midrule

\multicolumn{1}{l}{\multirow{2}{*}{$M_{500}>3.5 \times 10^{14} M_\odot$}}  & 1T & 0.36    &     0.038     &   3.7 (49./13)       \\
\multicolumn{1}{l}{}                                                     & 2T  & 0.35    &     0.043     &   1.5 (19./13)        \\ \bottomrule

\end{tabular}
\tablecomments{Similar to Table \ref{tab:zinstats}, using a cluster sample that excludes $f_{out}>0.25$ clusters.}
\end{center}
\end{table}

The second systematic effect we test is the effect of the number of radial bins used to measure $\bar{Z}_{in}$ in the obtained value. As can be seen in Equation \ref{eq:zavgin}, $\bar{Z}_{in}$ is a weighted sum of single metallicity measurements. For clusters without enough photons to create multiple radial bins within $0.15R_{500}$, the measurements will give less precise estimates of $\bar{Z}_{in}$, on average. We find, however, that there is no significant dependence of $\bar{Z}_{in}$ on the number of bins used to estimate it. In addition, we also find that the number of bins covering $r<0.15R_{500}$ does not depend on $M_{500}$.

We also repeat the analysis using a constant physical radius aperture of 150kpc to compute the inner iron mass fraction. We find that the different levels of dispersion remain unchanged, even with the physical radius aperture.

\subsection{Metallicity Scatter as a Reflection of Structure Formation}
\label{sec:zstructform}

One possible physical explanation to this observed difference between low-mass and high-mass clusters, is that the metal content in clusters is driven by the merger history of clusters. In the hierarchical model of structure formation, low-mass clusters, groups and galaxies merge to form the larger-mass clusters. Thus, if the metal content is non-uniform across all these progenitors, as they merge with each other, the resulting metallicity is an average of the initial progenitor metallicities. In a simple model, if the metallicities of these progenitor structures are distributed around a mean universal value, $Z_0$, then the sum metal content in a cluster formed by the merger of all these components should approach $Z_0$, as the number of components increases. The iron mass fraction, $\bar{Z}_{in}$, of a large cluster will thus be an average of the metallicities of its smaller progenitors.

Under this hypothesis, the decrease of the dispersion of $\bar{Z}_{in}$ as we go from low-mass clusters to high-mass clusters simply results from the mixing of low-mass clusters' gas, after they merge to make larger clusters. The mixing then results in averaged, less dispersed metallicity values in the merged clusters, compared to the initial progenitors' metallicities. It must be noted, however, that a high mass cluster from our sample, say of $M_{500} \sim 10^{15} M_\odot$, will not be exclusively formed by the merger of $10^{14} M_\odot$-sized clusters, i.e from the low-mass extreme of our sample. A $10^{15} M_\odot$ cluster will undergo numerous mergers involving galaxy- and group-sized haloes, as well as a smooth and continuous accretion \citep[see e.g.][]{fakho10}. This does not contradict the above hypothesis, as the more numerous the components making up a cluster are, the closer its metallicity approaches the universal average value.

If this is the correct explanation for the observed larger dispersion of $\bar{Z}_{in}$ in low-mass clusters, then it should also be reflected in the outer radii metallicity, $\bar{Z}_{mid}$, which is measured in the range $0.15<r<0.3R_{500}$. Our analysis however does not detect the same signal in the outer regions, as we do for inner metal content. The sample dispersion in $\bar{Z}_{mid}$ is roughly 0.079$Z_\odot$ (0.073$Z_\odot$) for the low-mass (high-mass) sample, and the contribution of measurement uncertainties to that dispersion is estimated to be around 0.04$Z_\odot$ (0.05$Z_\odot$) for the low-mass (high-mass) sample.

The decrease of the dispersion of \textit{inner} metallicity from small to large clusters requires that, as clusters merge, metals from the progenitor clusters are able to efficiently find their way to the center of the cluster, while avoiding mixing with gas in the outer regions of clusters. Interestingly, the hydrodynamical simulations of \citet{cora06} credit the infall of cold metal-rich clumps from large radii for the metal enrichment of the cluster central regions. A similar process is also one of the mechanisms evoked in \citet{milli11} to explain the existence of a peak in the observed radial profiles of SNCC's metal products, also observed in e.g. \citet{sande06}, \citet{simio09} and \citet{simio10}. Metal-rich gas from a small cluster, or from a galaxy, merging into a larger one can avoid mixing with the bulk ICM at large radii if its entropy is low enough to allow it to pierce through the outer ICM and reach the center of the cluster. In addition, the absense of this discrepancy in the outer regions of clusters could imply that metals in the outer regions come from a source with a more uniform metallicity level, i.e. a source whose metallicity varies less from cluster to cluster. Such a homogeneous source could be the gas that is accreted very early in the formation of clusters, and whose metal contribution is generally referred to as pre-enrichment, which we further discuss in Section \ref{preenrich}. Along the same lines, a slightly different interpretation of this finding is that, as time goes by, the haloes that merge later tend to have a wider distribution of metallicities, than those that merge earlier.

The hypothesis that merger statistics is behind the observed metallicity distribution across clusters can be tested in models that combine the statistics of structure formation with metal production in merging haloes, using a semi-analytical approach. Such a model was built in \citet{elkho12} to test other aspects of cluster chemical and dynamical histories, and can be adapted to test whether such observations can be reproduced semi-analytically.

We provide here, however, a very crude test of the above hypothesis. We would like to test whether the metallicities of clusters formed through the mergers of low-mass clusters from our sample have a similar distribution to that of metallicities of our high-mass clusters. To this end, we start by computing the metallicity $\bar{Z}_{in}$ resulting from the merger of two clusters from the sample of clusters in Figure \ref{fig:zin_m500} with $M_{500}<3.5 \times 10^{14} M_\odot$. We draw any two clusters from the low-mass subset, and define the metallicity resulting from their merger as the gas-mass--weighted average of the $\bar{Z}_{in}$ values of the two merging clusters. The gas mass used in weighting the average is that measured within $0.15R_{500}$, from the data. The resulting metallicity is assigned to a mass, $M_{500}$, which is the sum of the masses of the merging clusters. This is repeated with three-, four-, five- and six-cluster mergers, again based on our $\bar{Z}_{in}$ measurements in low-mass clusters. Our generated metallicity distribution thus comes from all possible mergers between 6 or less clusters from our low-mass clusters sample.  

\begin{figure}[h]
\centering
% plot generated by ~/caviar/research/acchif3/zdisp_fromdata2g2.pro
\scalebox{0.6}{\includegraphics*{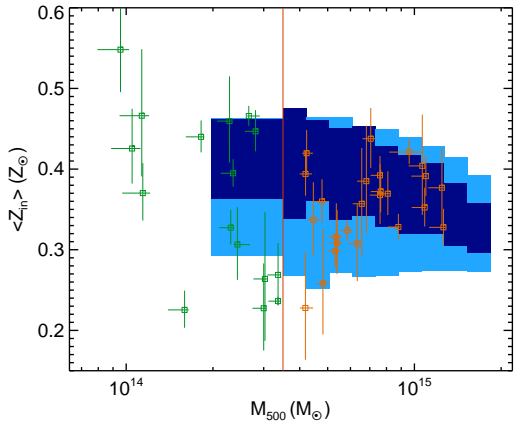}}
\caption{The width of the metallicity distribution, resulting from the hypothetical merger of low-mass clusters from our sample. The light (dark) blue region represents the 95\% (68\%) spread of generated metallicities. The green and orange data points represent our measurements in low- and high-mass clusters, respectively. The vertical line is plotted at $M_{500}=3.5 \times 10^{14} M_\odot$, and separates the above 2 samples.}
\label{fig:zindisp}
\end{figure}

Figure \ref{fig:zindisp} shows the result of the above simulation, superimposed on our data points. We recover qualitatively the trend in our data, whereby the width of the distribution of clusters' metallicities decreases with mass. Beyond this toy model, a model incorporating more detailed statistics of structure formation, as well as a more sophisticated model for metal content in merging clusters, is required to lend support to the hypothesis linking our metallicity measurements to structure formation.

\subsection{Support for Pre-Enrichment?}
\label{preenrich}

We argue above that the fact that metallicity dispersion as a function of cluster mass is not observed to change in the outer regions, while it does vary in inner regions can be explained if, outside the cluster core, most of the metals are the result of pre-enrichment, which is the metallicity level set before cluster formation \citep[e.g.][]{fujit08,matsu13,werne13}. The pre-enriched gas would have an approximately universal metallicity level, compared to the cold infalling haloes contributing metals at later times, whose gas metallicity values are more diverse.

We find another clue pointing to this initial pre-enrichment metallicity level --- presumed to be the same for all clusters --- when we compare $\bar{Z}_{in}$ values to the stellar-to-gas mass fractions in clusters. \citet{dai10} measure the stellar and baryon mass fractions of clusters with temperatures $1 \lesssim kT_X \lesssim 10$ keV, using 2MASS data for optical measurements, and \textit{ROSAT} data for X-ray measurements. They measure a decreasing stellar-to-gas mass ratio, $r_{sg}$, as a function of cluster temperature, $kT_X$. Using the results of \citet{dai10}, we can thus estimate the stellar mass from the X-ray temperature of a cluster and test whether it correlates with the metallicity of the ICM. A correlation is expected if the population of stars producing the metals detected in X-ray is the same as the one producing the optical luminosity of galaxies.

\citet{dai10} fit the temperature--stellar-mass trend to a power-law and obtain the following best-fit relation:

\begin{equation}
\log(r_{sg}) = -0.65-1.03 \log \left( \frac{kT_X}{1\mbox{keV}} \right) \ .
\end{equation}

Now, we recall that $\bar{Z}_{in}=M^{in}_{Fe}/(A_{Fe}M^{in}_{gas})$, where $M^{in}_{Fe}$ and $M^{in}_{gas}$ are the iron and gas masses interior to $r=0.15R_{500}$, respectively, and $A_{Fe}$ is the solar iron abundance. We can then write

\begin{equation}
\bar{Z}_{in} = \frac{M_* \langle M_{Fe}/M_*\rangle}{M^{in}_{gas} A_{Fe}} , \
\end{equation}

\noindent where $M_*$ is the stellar mass of the cluster, and $\langle M_{Fe}/M_*\rangle$, the average iron mass in the ICM per stellar mass. Or, defining $Z_{eff}=\langle M_{Fe}/M_*\rangle/A_{Fe}$, this becomes

\begin{equation}
\bar{Z}_{in} = Z_{eff} r_{sg} \left(kT_X \right) \ ,
\label{eq:zinrsgnocont}
\end{equation}

\noindent where $Z_{eff}$ is the average iron mass in the ICM per stellar mass, scaled by the solar iron abundance, $A_{Fe}$. The assumption here is that $Z_{eff}$ will be the same for all clusters, and will not depend on $kT_X$.

If we assume that all the iron inside $0.15R_{500}$ has been produced by the stars in the galaxies whose luminosities were used by \citet{dai10} to measure $r_{sg}$, above, or that at least a population of stars of mass proportional to $M_*$ produced all the iron observed, then Equation \ref{eq:zinrsgnocont} should describe the $\bar{Z}_{in}-kT_X$ data, once scaled by a suitable $Z_{eff}$.

\begin{figure}[h]
\centering
% plot generated by ~/caviar/research/acchif3/z_m500_ktx_mid2nozbin_allclu2.pro
\scalebox{0.5}{\includegraphics*{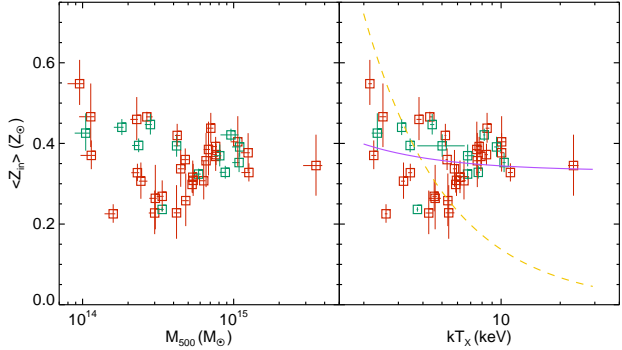}}
\caption{\textit{Left:} $\bar{Z}_{in}$ vs. $M_{500}$. \textit{Right:} $\bar{Z}_{in}$ vs. $kT_X$. Red points are for NCC clusters, while turquoise points are for CC clusters. The yellow dashed line represent the $\bar{Z}_{in}-kT_X$ best-fit based on the stellar-to-gas mass ratio. The purple line is a similar fit based on stellar-to-gas mass ratio and assuming in addition a baseline pre-enrichment level.}
\label{fig:zin_m500str}
\end{figure}

We perform a least-squares fit of Equation \ref{eq:zinrsgnocont} to the data. The purpose of this fit is not to estimate $Z_{eff}$, but simply to test whether the above picture is consistent with our data. We assume equal errors on metallicity measurements so that the results are not largely biased by the datapoints with smaller uncertainties. The best-fit model plotted in Figure \ref{fig:zin_m500str}, as the dashed yellow line, shows that Equation \ref{eq:zinrsgnocont} is an inadequate fit to the observations. Similar to the conclusion in \citet{bregm10}, this suggests that the stellar population producing the metals within $r=0.15R_{500}$ is unlikely to be related to the currently observed galaxy population in clusters.

At the risk of having a model too flexible for our dataset, we thus introduce another parameter into Equation \ref{eq:zinrsgnocont}, which is the mean metallicity in the ICM believed to have been set before the current galaxy population started adding more metals. We denote this initial metallicity by $Z_{pre}$. Our model then becomes

\begin{equation}
\bar{Z}_{in} = Z_{eff} r_{sg} \left(kT_X \right) + Z_{pre} \ .
\label{eq:zinrsg}
\end{equation}

Unlike Equation \ref{eq:zinrsgnocont}, Equation \ref{eq:zinrsg} fits the high-mass data better \textit{and} passes near the middle of the wide distribution of metallicities of low-mass clusters. The best-fit is shown in Figure \ref{fig:zin_m500str} as the purple line. This gives support to a mass-independent initial metallicity level in the cores of clusters. In this picture, larger clusters would have most of their core metals set by pre-enrichment, as $Z_{pre} \gg Z_{eff} r_{sg}$ for large $kT_X$. Conversely, smaller clusters would have a larger contribution from metals associated with stars, as $Z_{eff} r_{sg} \left(kT_X \right)$ increases at low $kT_X$. The best-fit model has a pre-enrichment value of $Z_{pre} \approx 0.3Z_\odot$, and a $Z_{eff} \approx 2. Z_\odot$. While the value of $Z_{pre}$ agrees with expectations, the large $Z_{eff}$ value again points to the lack of observed galaxies, compared to the observed metals in clusters. The two fits above give similar results when repeated after excluding the bias-suspected clusters, i.e. clusters with $f_{out}<0.25$.

\subsection{Inner Metallicity Boost During Mergers}

Finally, we test whether the dispersion in $\bar{Z}_{in}$ is related to dynamic activity, as measured by the centroid shift, $\langle w \rangle$, which is the size of the scatter of the X-ray centroid measured within various apertures around the X-ray peak \citep[e.g.][]{mohr93,ohara06,poole06}, and which is ideal for capturing merger activity \citep{poole06}. We find weak evidence for a $\bar{Z}_{in}-\langle w \rangle$ correlation, as we describe below.

\begin{figure}[ht]
%\centering
% plot generated by ~/caviar/research/acchif3/sz6_w3_zout_ac3b2.pro
\scalebox{0.5}{\includegraphics{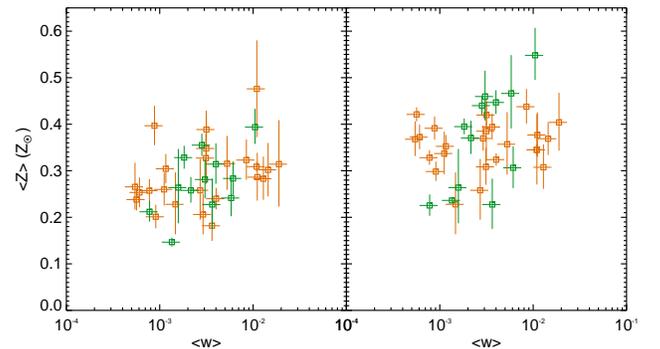}}
\caption{\textit{Left:} $\bar{Z}_{mid}$ vs. $\langle w \rangle$. \textit{Right:} $\bar{Z}_{in}$ vs. $\langle w \rangle$. Green data points in either panel are for low-mass clusters, where $M_{500} < 3.5 \times 10^{14} M_\odot$. Orange data points are for $M_{500}>3.5 \times 10^{14} M_\odot$ clusters.}
\label{fig:zvsw}
\end{figure}

Following the prescription in \citet{poole06}, we calculate the position of the centroid of the X-ray emission within a radius of $0.3R_{500}$, and excluding the central 30kpc. We then calculate the centroid for apertures that are successively smaller by 5\% of $0.3R_{500}$. For each aperture, $i$, we record the distance between calculated centroid and the X-ray peak, $d_i$. The centroid shift, $\langle w \rangle$, is then simply the standard deviation of the distances, $d_i$, scaled by $R_{500}$. We compute $\langle w \rangle$ for clusters with a FOV that covers $0.3R_{500}$ entirely. We find this radius to be a good compromise to include a large area and a good number of clusters. We show the computed $\langle w \rangle$ values in Table \ref{tab:morph}. Figure \ref{fig:zvsw} shows a plot of $\bar{Z}_{in}$ vs. $\langle w \rangle$, in the right panel. We also include the same plot for $\bar{Z}_{mid}$, in the left panel of the figure. Datapoints corresponding to low-mass clusters are in green color to distinguish them from the high-mass cluster datapoints in orange.

Figure \ref{fig:zvsw} suggests that there might be a correlation between $\bar{Z}_{in}$ and $\langle w \rangle$, for low-mass clusters. Pearson's product-moment correlation coefficient is found to be $r_P=0.66$, with a p-value of 1.0\% for the $\bar{Z}_{in}-\langle w \rangle$ correlation in clusters with $M_{500} < 3.5 \times 10^{14} M_\odot$.

This result suggests an alternative explanation to the $\bar{Z}_{in}$ dispersion in low-mass clusters, whereby mergers boost measured metallicity. This can be achieved in one of two ways. First, if mergers induce central AGN activity, then we might be observing metals distributed into the central cluster region by the central active engine, as was observed and described in e.g. \citet{simio09,simio09b,kirk09,kirk11,osull11}. This can also be the reason that we observe correlations between $\bar{Z}_{in}$ and the scaled entropy, $x_s$, as was shown in Section \ref{sec:zxs}. Second, this enhanced central metallicity might simply be the measurement of the central metallicity peak of an infalling sub-cluster that is not completely merged with the main cluster, and whose emission is superimposed on the $r<0.15R_{500}$ region of the main cluster. In both cases, these effects would need to have a stronger influence on metallicity in low-mass clusters compared to larger clusters.

\subsection{Searching for a Pre-Enrichment--Pre-Heating Link}

The motivation behind undertaking the study of metallicity and entropy as far as possible from the central region --- where the influence of the central engine and the effects of cooling increase --- was to search for the signature of supernovae (SNe), which have been heating and chemically enriching the gas surrounding them even before the formation of the galaxy clusters.

The plots of $\bar{Z}_{mid}$ vs. $x_s$ in Section \ref{sec:zxs} show no hint of a relation between the two quantities. However, we note that for a given radius, the range of values for $x_s=S/S_{grav}$ can span a range as large as a factor of 3, at large radii. One can thus envision further study at such large radii to be applied to a larger sample of clusters, which is available in current \textit{Chandra} archival data. A stacking technique can be used to look for trends between metallicity and excess entropy, and to lower any systematics due to the high X-ray background count fraction at such high radii. For example, we could group clusters in bins of $S/S_{grav}$, where $S$ is measured at a large scaled radius, then extract spectra from each group of clusters from uniform $R_{500}$-scaled radial bins. These spectra could then be simultaneously fit, assuming they all have the same metallicity at a given scaled radial bin, and allowing for the temperatures to vary to match each cluster's temperature. Such study could be more sensitive to a potential weak trend between metallicity and excess entropy, pointing to the effects of early supernova enrichment and heat injection.

\section{Conclusion}

In this work, we analyze a sample of 46 galaxy clusters, extracting chemical and dynamical measurements, in the hopes of obtaining clues about the history of clusters. We measure entropy profiles out to the largest radii where temperature can be measured, and provide the best-fit temperature and density profiles for the community to use. We also measure metallicity profiles, for our cluster sample, and present them below. The data is made available on an FTP site\footnote{\texttt{ftp://space.mit.edu/pub/tamer/ebc2015/}}.

We observe a difference in the scaled iron mass between the centers of low-mass clusters, and the centers of high-mass clusters: the values of the iron content in small clusters are more dispersed than those in large clusters. We suggest two possible interpretations of this observation:

\begin{enumerate}

\item
The lower dispersion in the larger clusters may be a result of the averaging of metallicities from the larger number of haloes that have merged to form them. The fact that this effect can be seen even in the core of clusters lends support to the idea that the centers of clusters continue being enriched by cold and metal-rich gas, originating from the cluster outskirts, even at low redshift.

\item
Alternatively, there are hints that clusters can undergo a boost of metallicity during a merger event, which can contribute to the enhancement of metallicity measured in low-mass clusters. 

\end{enumerate}

We also look for a connection between the bulk metal content of clusters and their dynamical state, as measured by the deviation of their entropy profiles from a self-similar profile, expected from gravitational shock heating, during cluster formation. We find no evidence of such relation in our data. More sophisticated studies using a larger sample would be required for such measurement to obtain more conclusive results.

\acknowledgements
TYE would like to thank Michael McDonald for very fruitful conversations.

%%%%%%%%%%%%%%%%%%%%%%%%%%%%%% Appendix %%%%%%%%%%%%%%%%%%%%%%%%%%%%%%%%%%%%%
%%%%%%%%%%%%%%%%%%%%%%%%%%%%%%%%%%%%%%%%%%%%%%%%%%%%%%%%%%%%%%%%%%%%%%%%%%%%%
\clearpage

\appendix

\section{Entropy Profiles}
\label{app:sofr}

% plot again with different order so that we can see accept data
\begin{figure}[htp]
\centering
% plot generated by /home/tamer/research/acchif2/scatterkt7plotent.py
\includegraphics*[0.3in,0.4in][6.3in,6.4in]{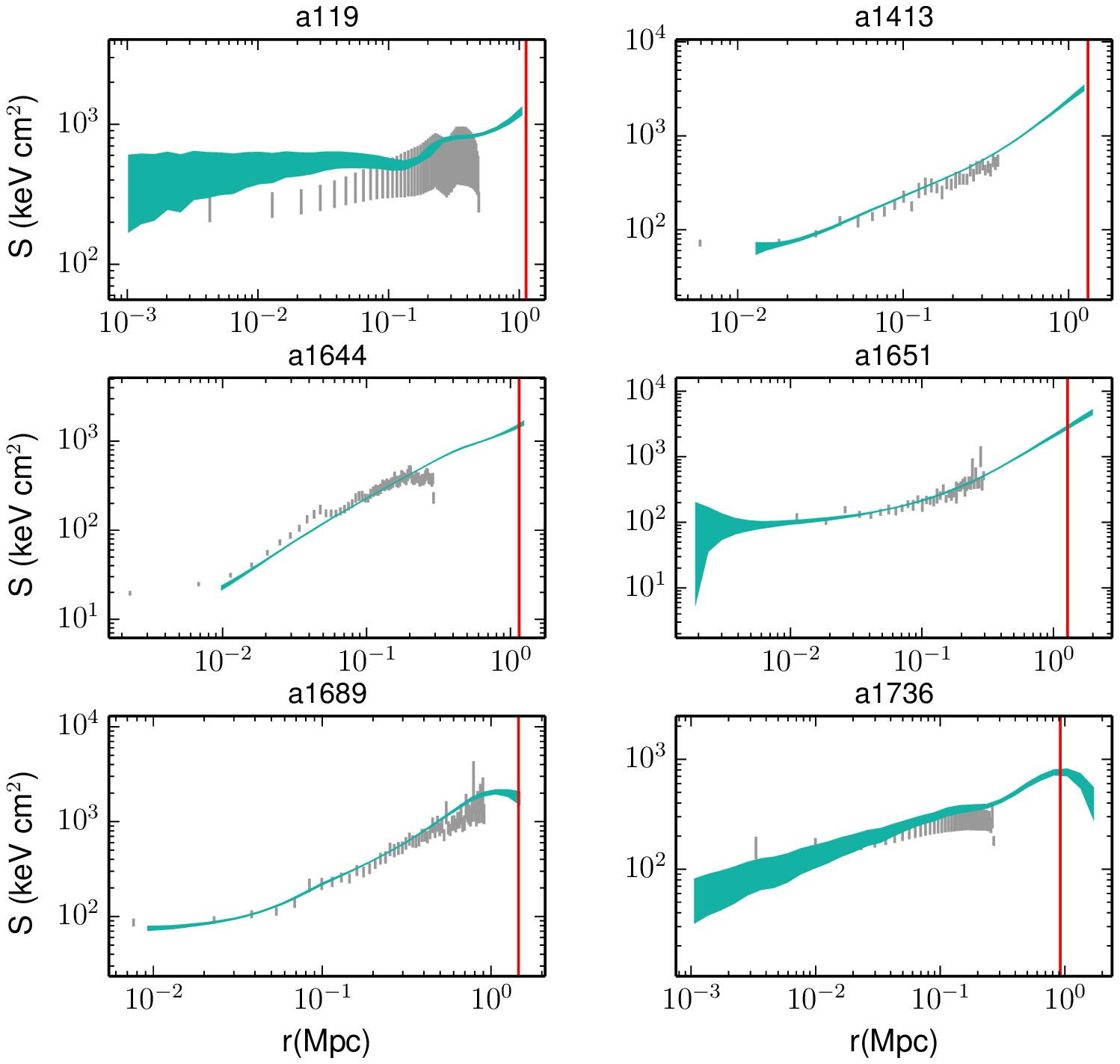}
\caption{Entropy profiles of the individual clusters in the sample. The turquoise area represents the 68\% confidence region computed at various radii, from the temperature and density profiles produced by the Monte Carlo simulations described in Section \ref{sec:uncert}. The gray error bars are the entropy profile measurements of the ACCEPT study, \citep{cavag09}. See text for details. The vertical red line is drawn at $r=R_{500}$.}
\label{fig:sofr}
\end{figure}

\begin{figure}[htp]
\centering
\ContinuedFloat
% plot generated by /home/tamer/research/acchif2/scatterkt7plotent.py
\includegraphics*[0.3in,0.3in][6.3in,8.3in]{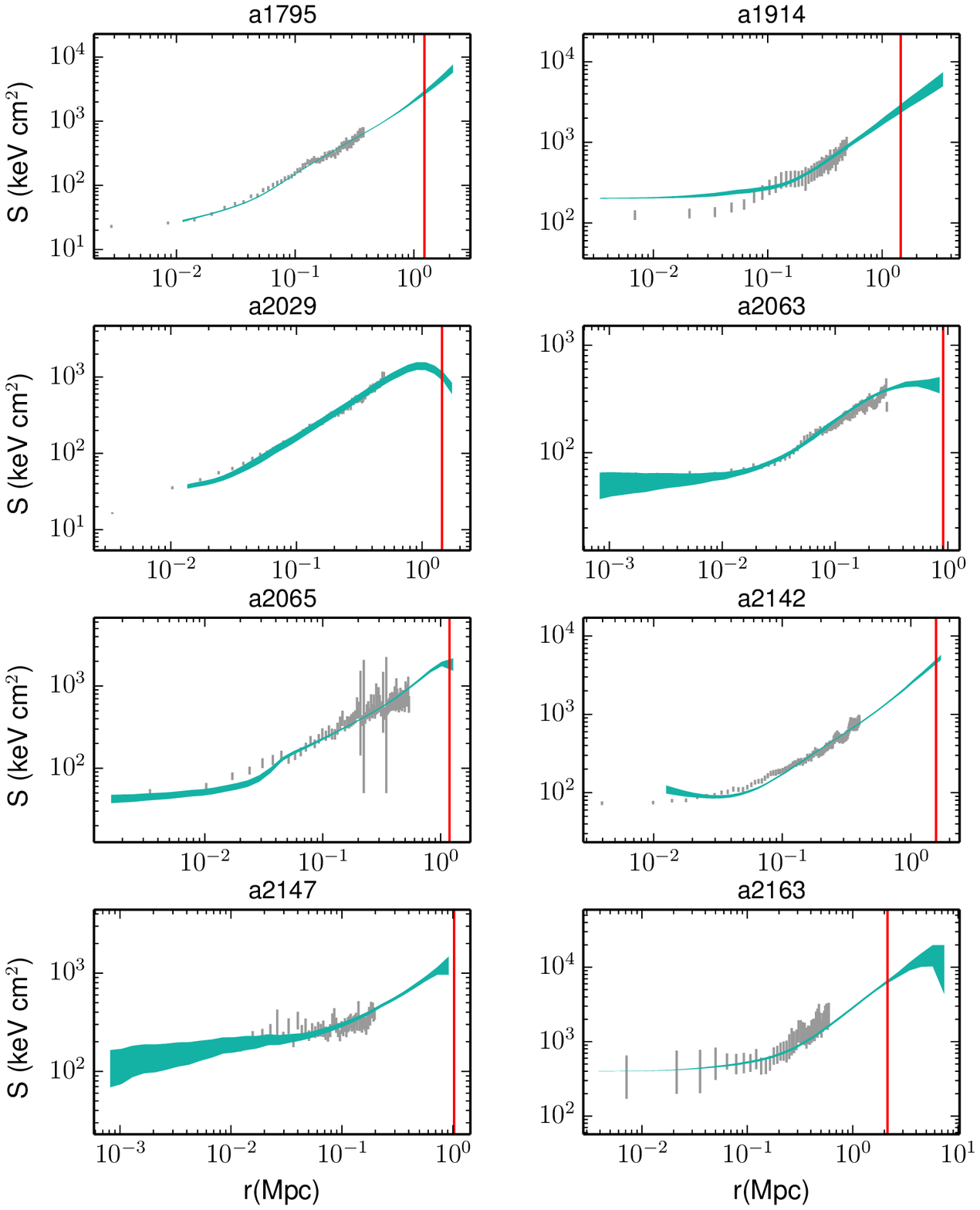}
\caption[]{\textit{Continued.}}
%\label{fig:sofr}
\end{figure}

\begin{figure}[htp]
\centering
\ContinuedFloat
% plot generated by /home/tamer/research/acchif2/scatterkt7plotent.py
\includegraphics*[0.3in,0.3in][6.3in,8.3in]{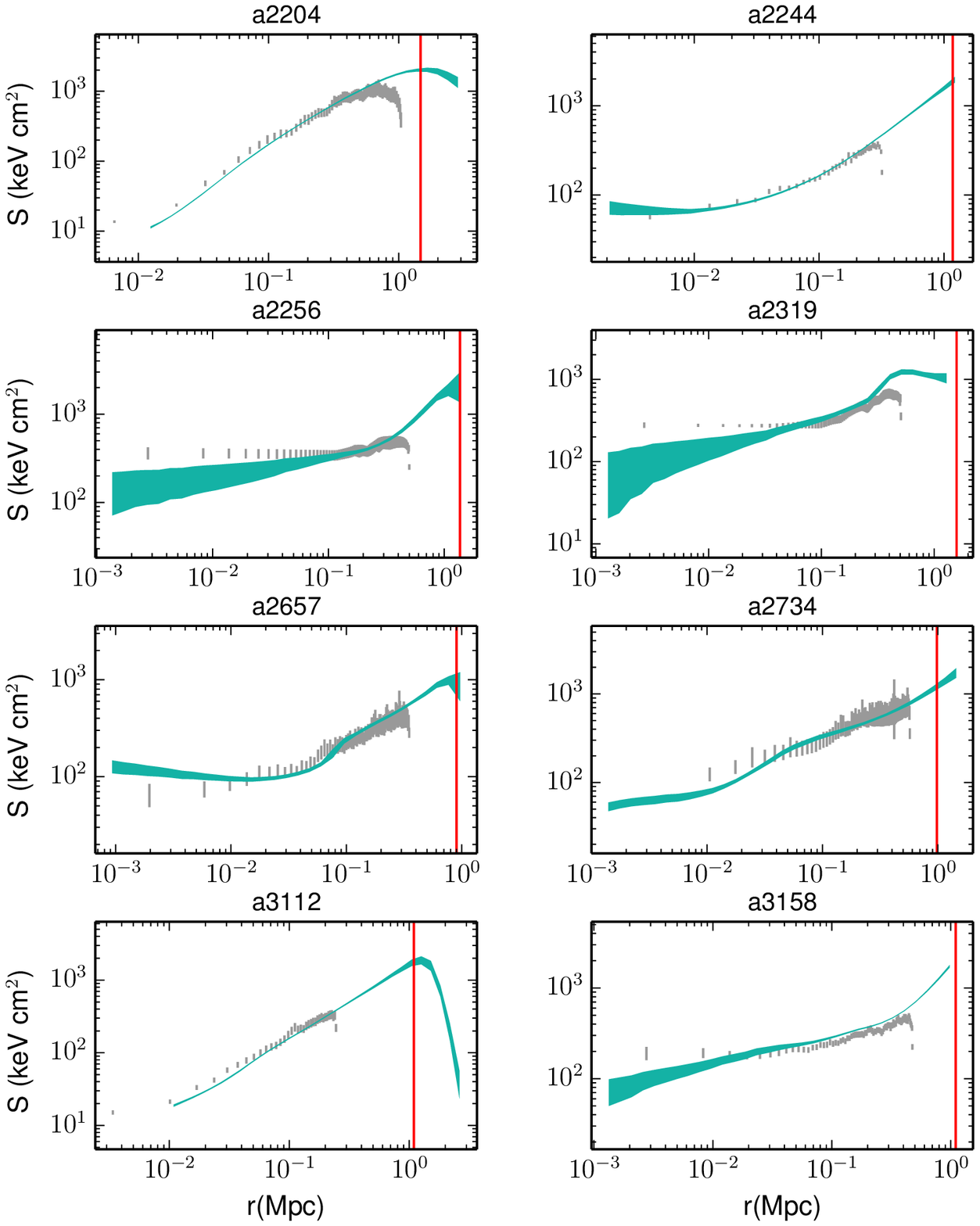}
\caption[]{\textit{Continued.}}
%\label{fig:sofr}
\end{figure}

\begin{figure}[htp]
\centering
\ContinuedFloat
% plot generated by /home/tamer/research/acchif2/scatterkt7plotent.py
\includegraphics*[0.3in,0.3in][6.3in,8.3in]{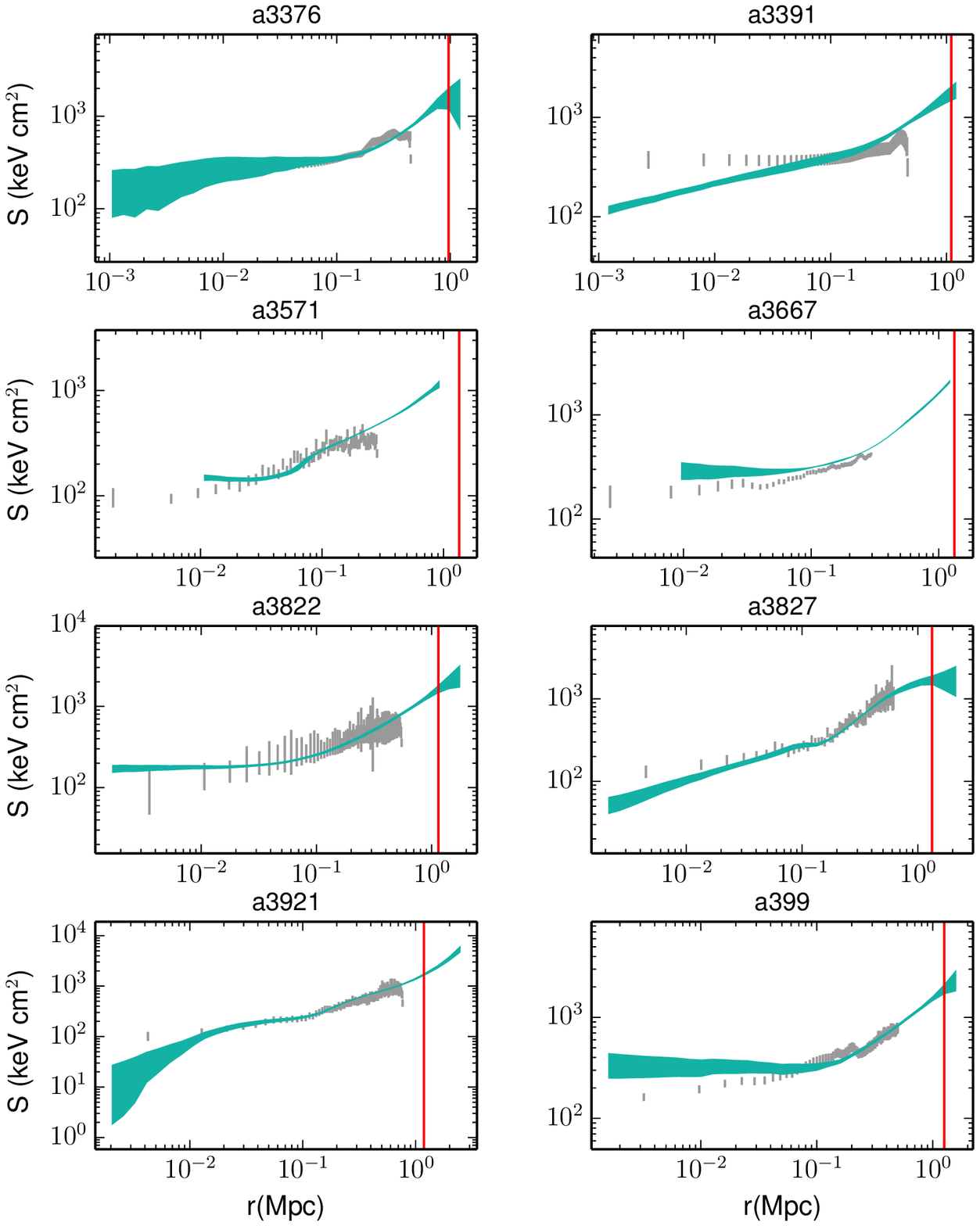}
\caption[]{\textit{Continued.}}
%\label{fig:sofr}
\end{figure}

\begin{figure}[htp]
\centering
\ContinuedFloat
% plot generated by /home/tamer/research/acchif2/scatterkt7plotent.py
\includegraphics*[0.3in,0.3in][6.3in,8.3in]{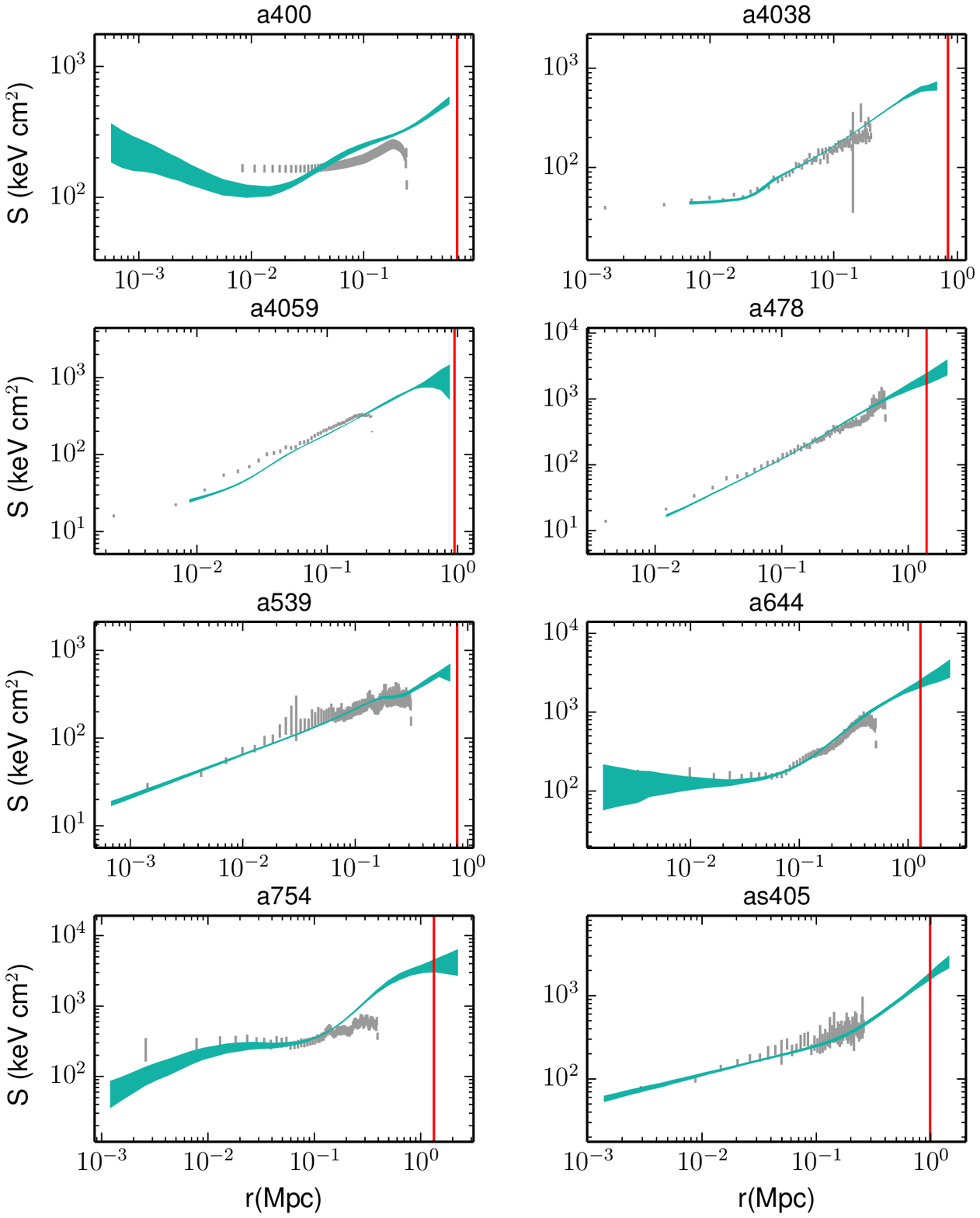}
\caption[]{\textit{Continued.}}
%\label{fig:sofr}
\end{figure}

\begin{figure}[htp]
\centering
\ContinuedFloat
% plot generated by /home/tamer/research/acchif2/scatterkt7plotent.py
\includegraphics*[0.3in,0.3in][6.3in,8.3in]{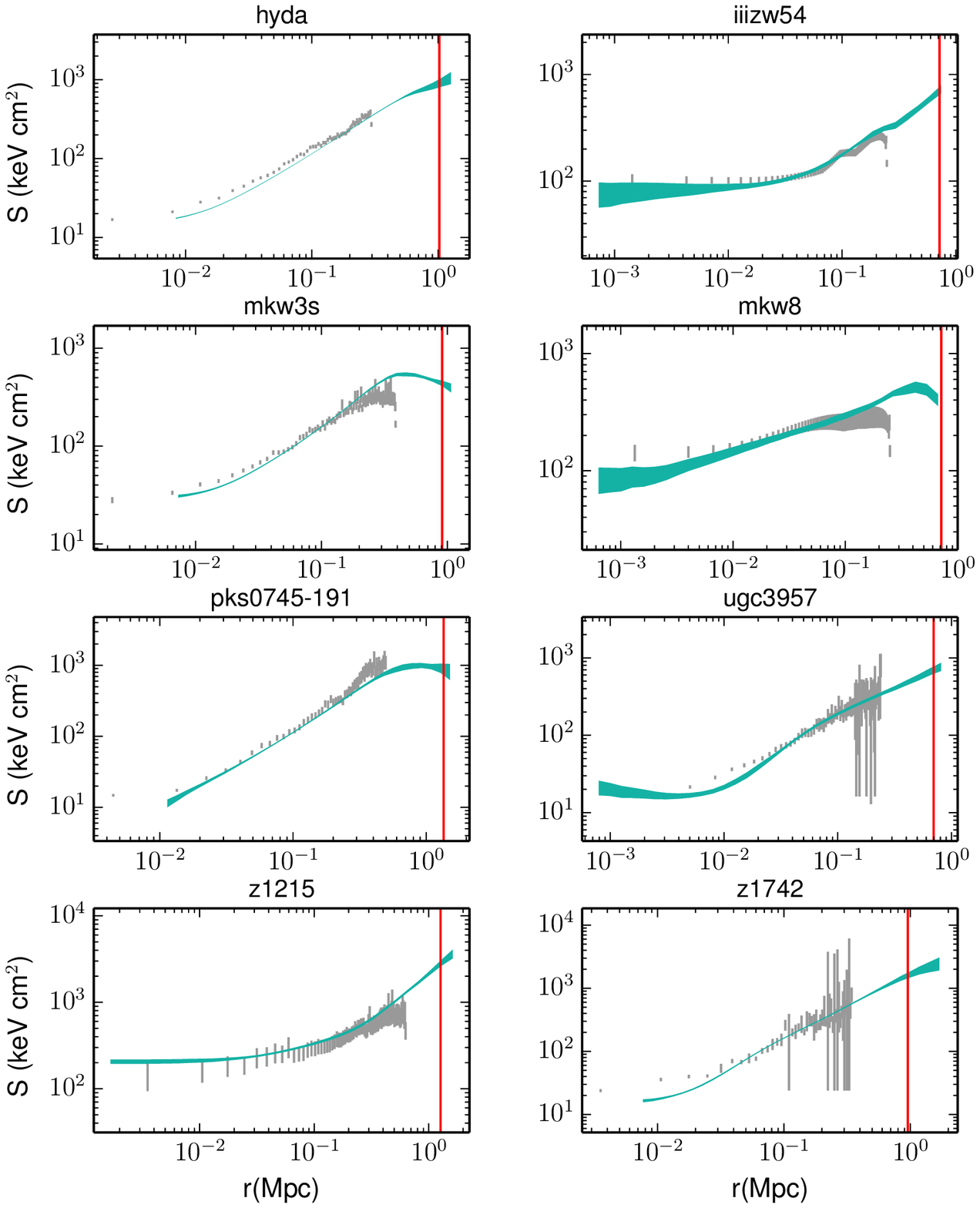}
\caption[]{\textit{Continued.}}
%\label{fig:sofr}
\end{figure}

%%%%%%%%%%%%%%%%%%%%%% Z(r) %%%%%%%%%%%%%%%%%%%%%%%%%%%%%%%%%%%%%%
\clearpage

\section{Metallicity Profiles}
\label{app:zofr}

\begin{figure}[htp]
\centering
% plot generated by ~/caviar/research/acchif3/plotzofr_sz10.pro
\scalebox{0.8}{\includegraphics*[0.15in,0.05in][8in,10in]{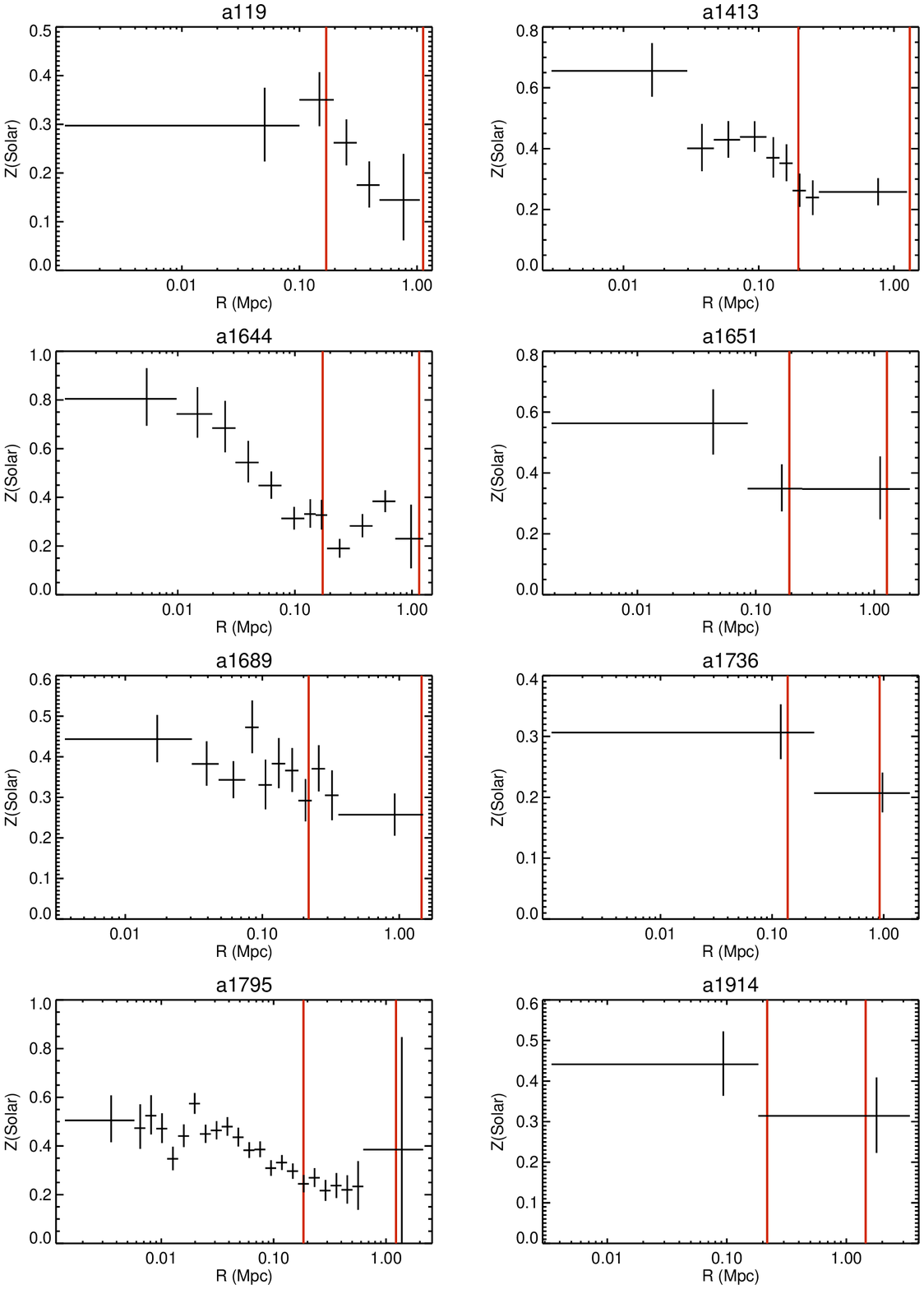}}
\caption[Individual cluster metallicity profiles]{Metallicity profiles of observed \textit{Chandra} clusters. The two vertical red lines are drawn at $r=0.15R_{500}$ and $r=R_{500}$. See Section \ref{sec:zmethod} for the details of the metallicity measurement method.}
\label{fig:zofr}
\end{figure}

\begin{figure}[htp]
\centering
\ContinuedFloat
% plot generated by ~/caviar/research/acchif3/plotzofr_sz10.pro
\scalebox{0.8}{\includegraphics*[0.15in,0.05in][8in,10in]{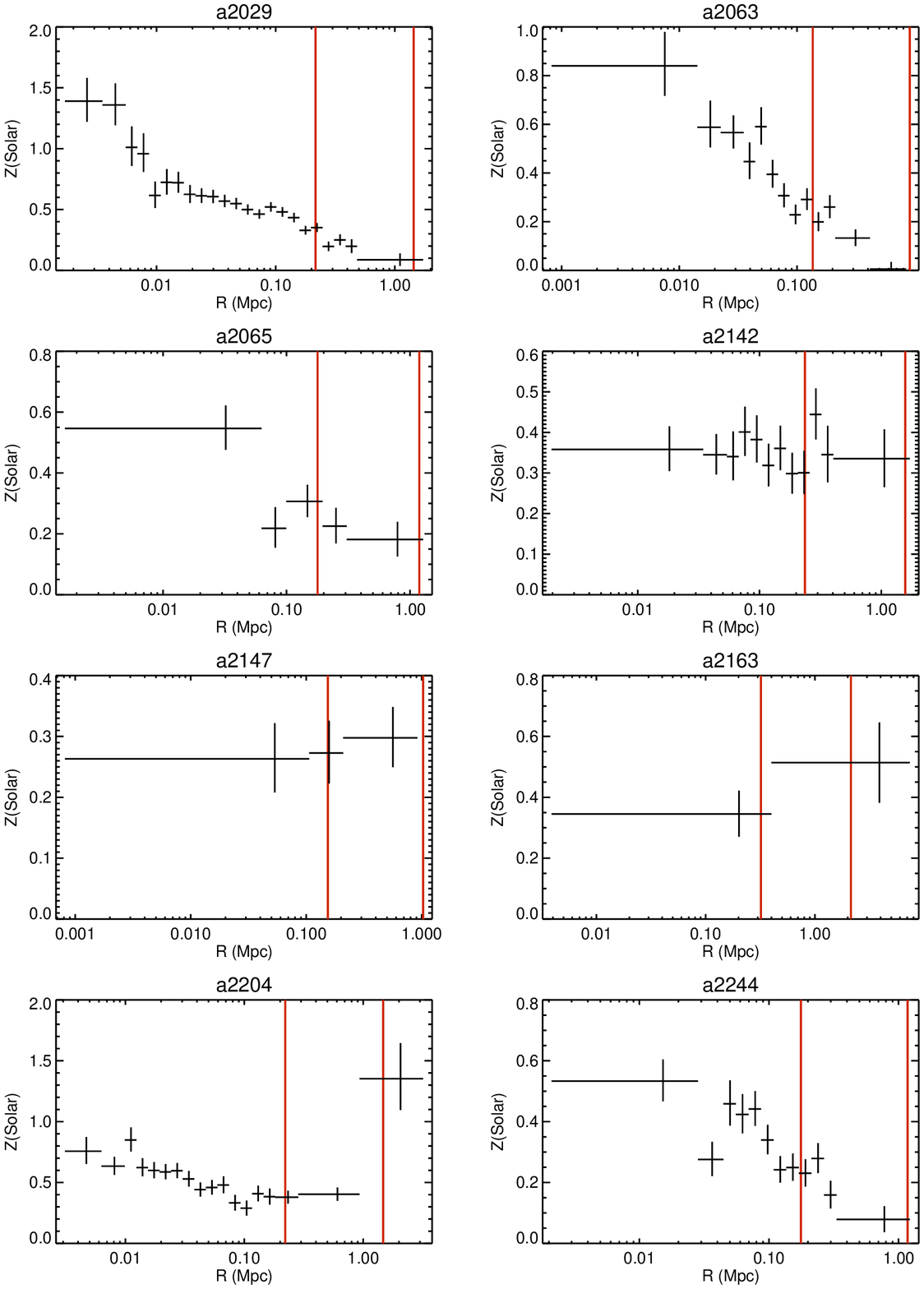}}
\caption[]{\textit{Continued.}}
%\label{fig:zofr}
\end{figure}

\begin{figure}[htp]
\centering
\ContinuedFloat
% plot generated by ~/caviar/research/acchif3/plotzofr_sz10.pro
\scalebox{0.8}{\includegraphics*[0.15in,0.05in][8in,10in]{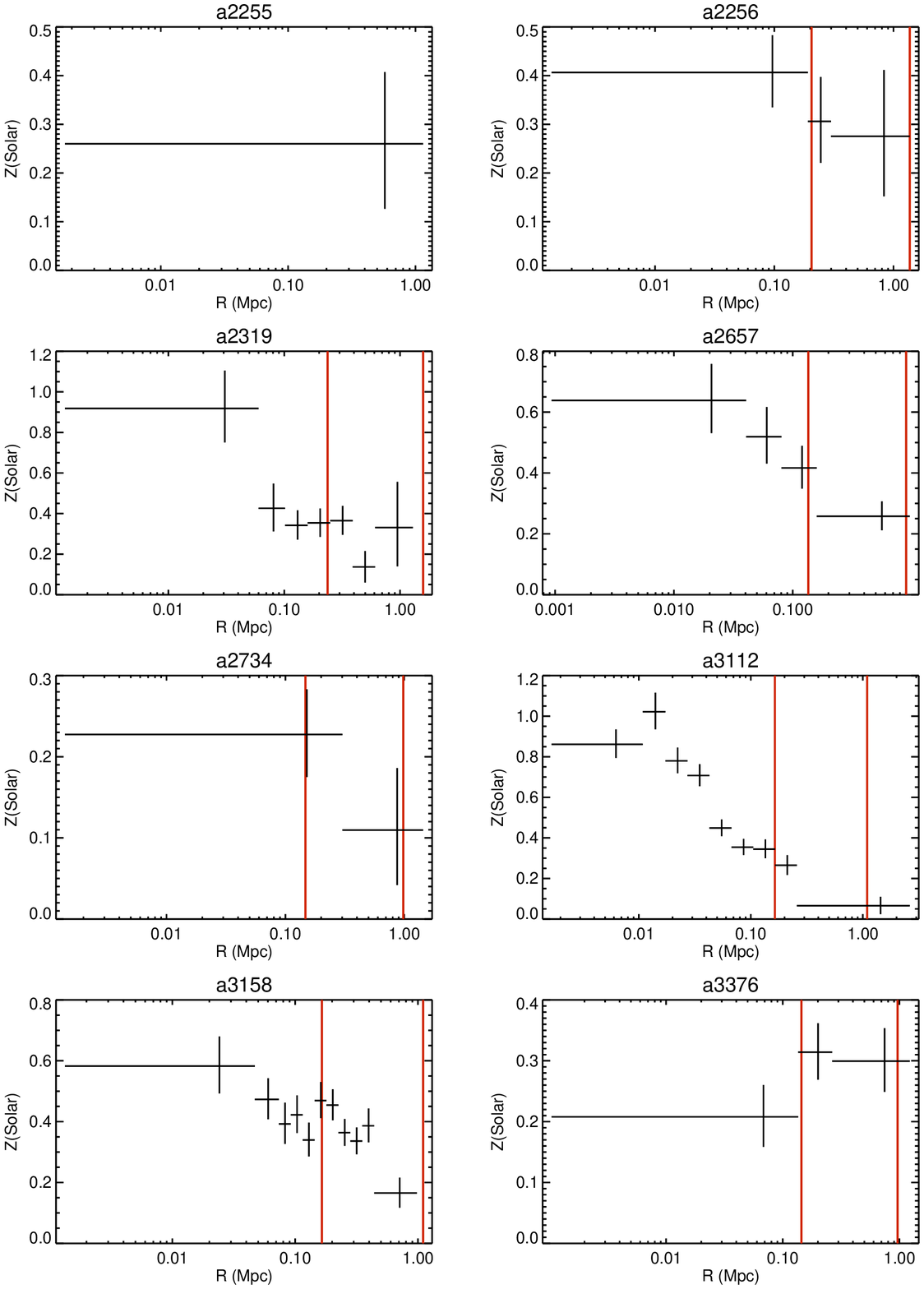}}
\caption[]{\textit{Continued.}}
%\label{fig:zofr}
\end{figure}

\begin{figure}[htp]
\centering
\ContinuedFloat
% plot generated by ~/caviar/research/acchif3/plotzofr_sz10.pro
\scalebox{0.8}{\includegraphics*[0.15in,0.05in][8in,10in]{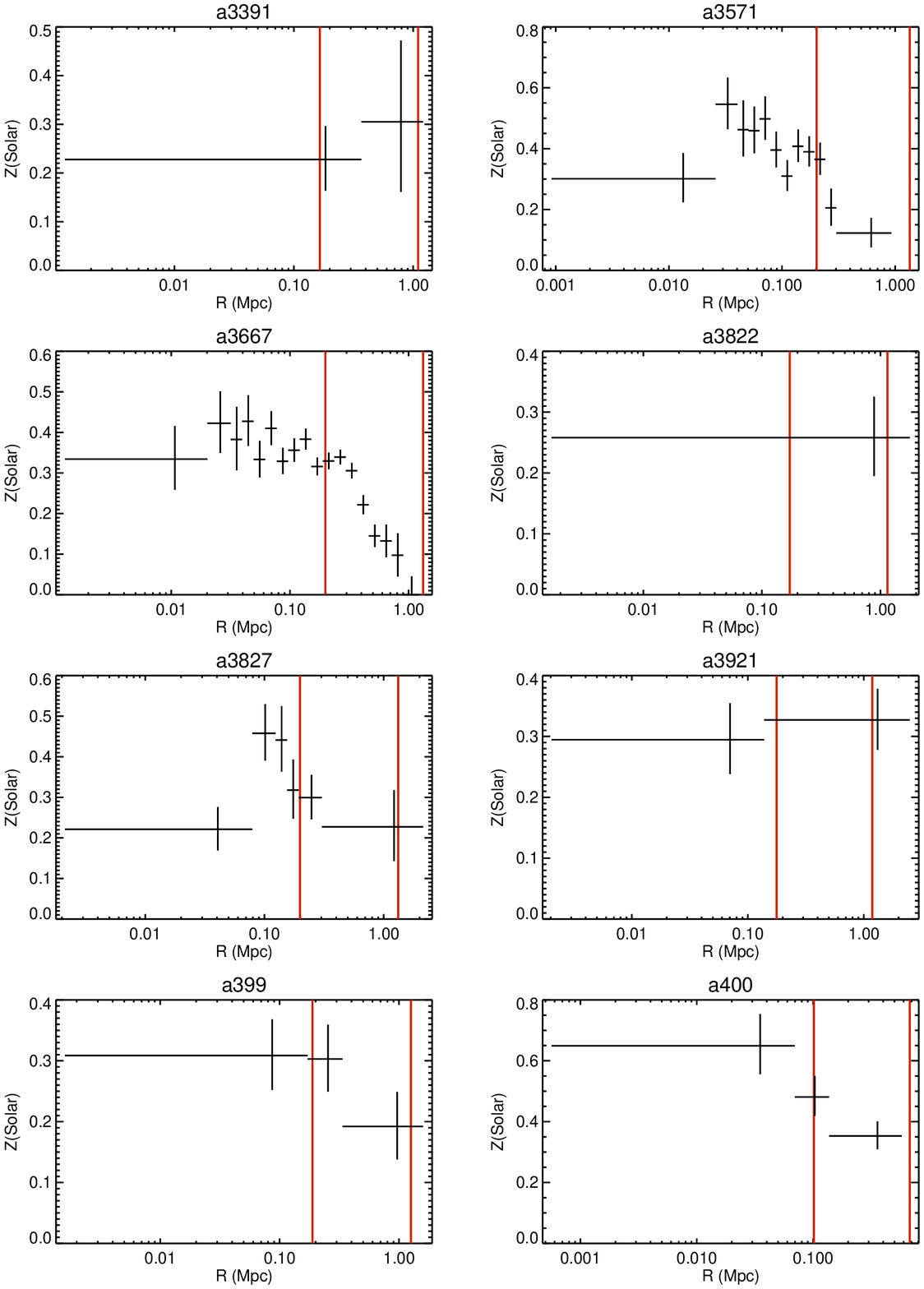}}
\caption[]{\textit{Continued.}}
%\label{fig:zofr}
\end{figure}

\begin{figure}[htp]
\centering
\ContinuedFloat
% plot generated by ~/caviar/research/acchif3/plotzofr_sz10.pro
\scalebox{0.8}{\includegraphics*[0.15in,0.05in][8in,10in]{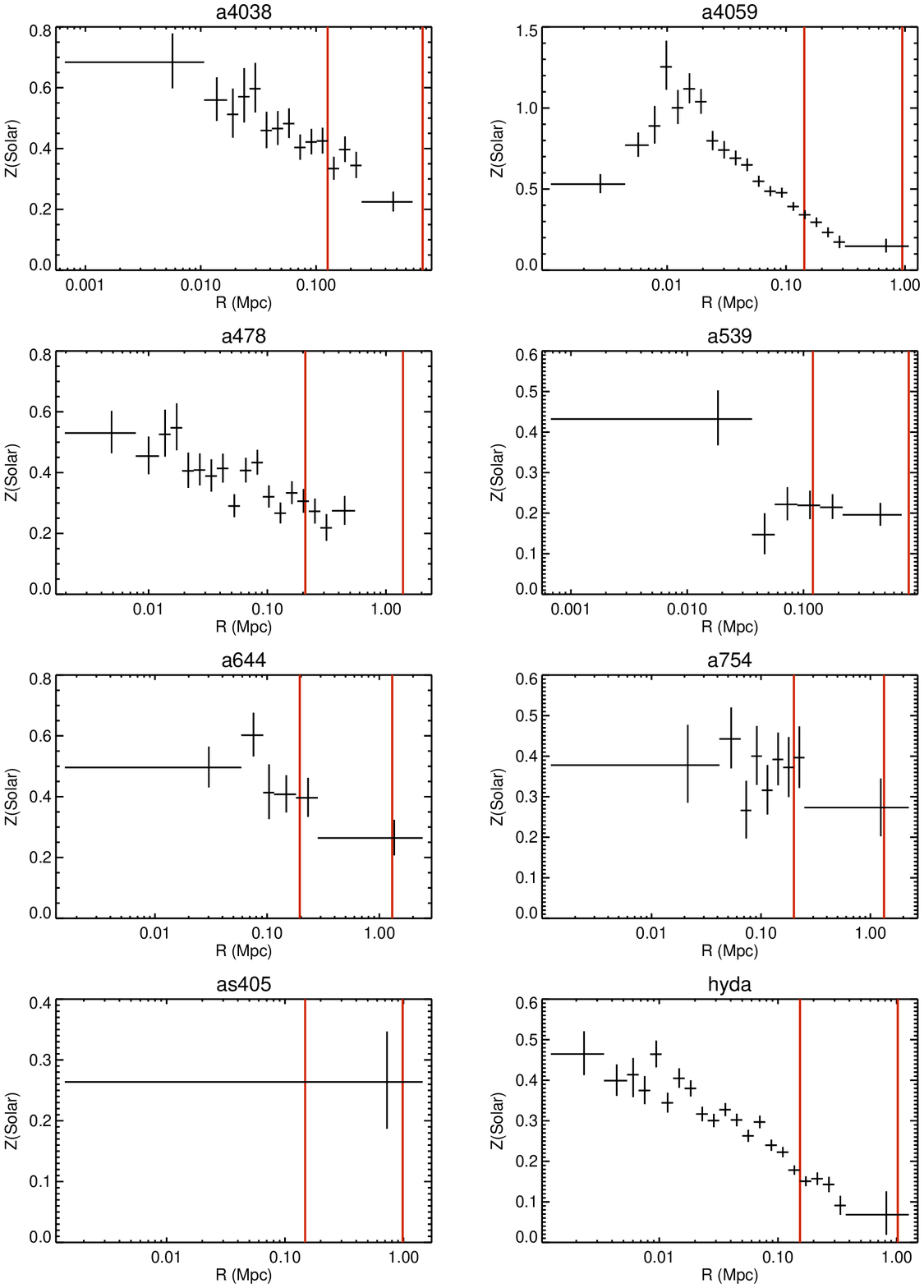}}
\caption[]{\textit{Continued.}}
%\label{fig:zofr}
\end{figure}

\begin{figure}[htp]
\centering
\ContinuedFloat
% plot generated by ~/caviar/research/acchif3/plotzofr_sz10.pro
\scalebox{0.8}{\includegraphics*[0.15in,0.05in][8in,10in]{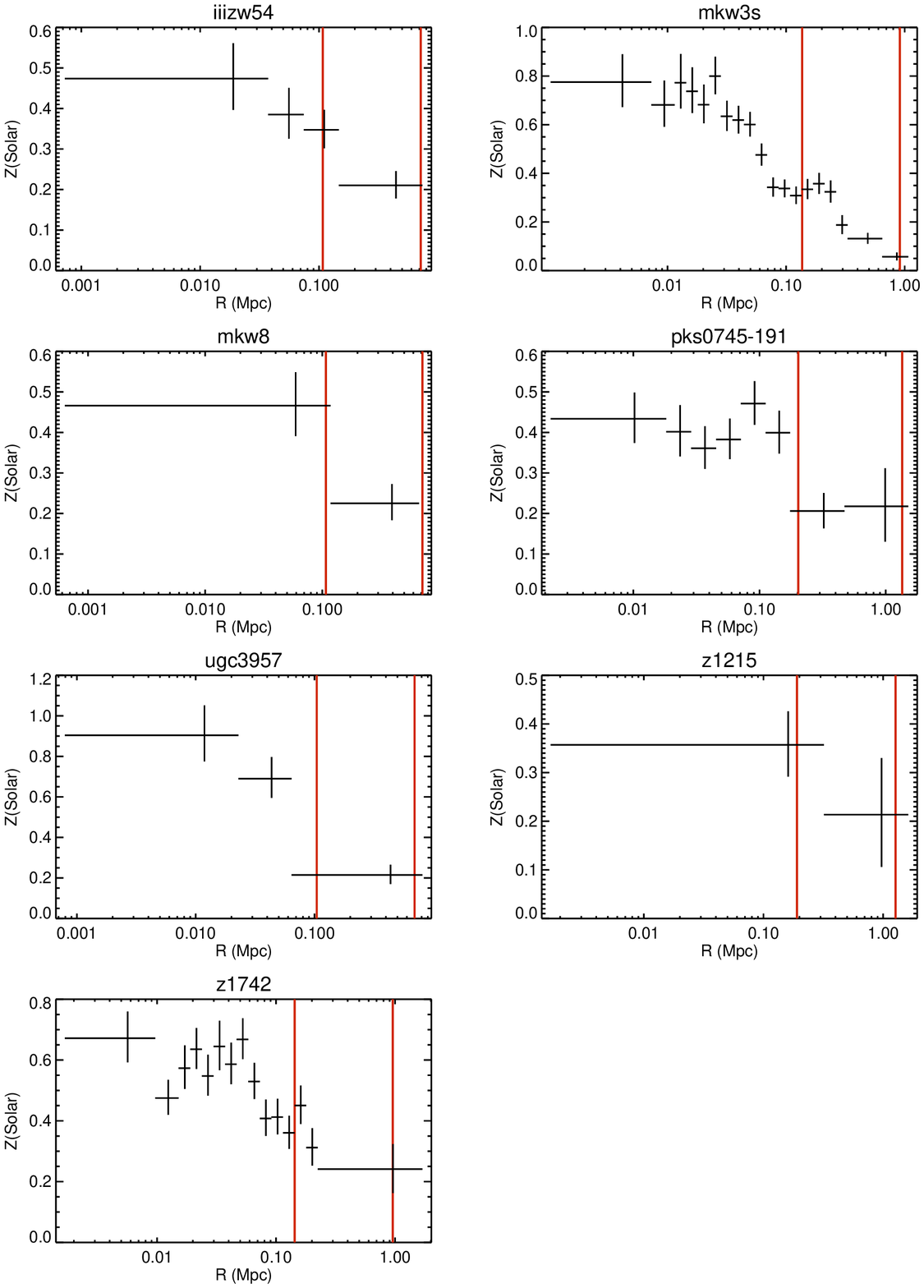}}
\caption[]{\textit{Continued.}}
%\label{fig:zofr}
\end{figure}

%%%%%%%%%%%%%%% M500, Mgas, R500, kTx %%%%%%%%%%%%%%%%%%

\clearpage
\section{Cluster Masses and Scales}
\label{app:masses}

\begin{table}[h]
\begin{center}
\caption{Cluster Masses and Scales.\label{tab:masses}}
\begin{tabular}{ @{}lllll@{} }
% updated to table_mass_ac3_pr.tex on 2/16/15
\toprule
Cluster & $kT_X$(keV) & $R_{500}$(Mpc) & $M_{500}(10^{14}M_\odot)$ & $ M_{gas}(10^{13}M_\odot)$ \\
\midrule
Abell 119 & $5.79^{+0.1}_{-0.1}$ & $1.13^{+0.02}_{-0.02}$ & $4.46^{+0.2}_{-0.2}$ & $4.50^{+0.1}_{-0.1}$ \\
Abell 1413 & $8.42^{+0.1}_{-0.2}$ & $1.31^{+0.007}_{-0.03}$ & $7.64^{+0.1}_{-0.6}$ & $8.11^{+0.1}_{-0.2}$ \\
Abell 1644 & $5.31^{+0.05}_{-0.09}$ & $1.15^{+0.02}_{-0.02}$ & $4.78^{+0.3}_{-0.3}$ & $5.50^{+0.2}_{-0.3}$ \\
Abell 1651 & $7.52^{+0.4}_{-0.3}$ & $1.28^{+0.01}_{-0.03}$ & $6.81^{+0.2}_{-0.5}$ & $7.30^{+0.2}_{-0.2}$ \\
Abell 1689 & $10.4^{+0.2}_{-0.2}$ & $1.45^{+0.005}_{-0.05}$ & $10.9^{+0.1}_{-1.}$ & $12.4^{+0.1}_{-0.4}$ \\
Abell 1736 & $3.19^{+0.05}_{-0.09}$ & $0.920^{+0.03}_{-0.009}$ & $2.43^{+0.3}_{-0.07}$ & $2.80^{+0.2}_{-0.1}$ \\
Abell 1795 & $6.75^{+0.05}_{-0.05}$ & $1.23^{+0.01}_{-0.02}$ & $5.86^{+0.2}_{-0.3}$ & $6.19^{+0.1}_{-0.1}$ \\
Abell 1914 & $10.1^{+0.5}_{-0.5}$ & $1.45^{+0.004}_{-0.05}$ & $10.7^{+0.1}_{-1.}$ & $12.3^{+0.2}_{-0.5}$ \\
Abell 2029 & $8.20^{+0.09}_{-0.09}$ & $1.44^{+0.06}_{-0.08}$ & $9.60^{+1.}_{-1.}$ & $12.5^{+2.}_{-3.}$ \\
Abell 2063 & $3.45^{+0.08}_{-0.09}$ & $0.906^{+0.03}_{-0.009}$ & $2.31^{+0.07}_{-0.2}$ & $2.36^{+0.07}_{-0.06}$ \\
Abell 2065 & $6.18^{+0.2}_{-0.2}$ & $1.19^{+0.02}_{-0.02}$ & $5.36^{+0.2}_{-0.3}$ & $5.80^{+0.1}_{-0.1}$ \\
Abell 2142 & $11.2^{+0.3}_{-0.3}$ & $1.57^{+0.01}_{-0.06}$ & $12.6^{+0.2}_{-1.}$ & $14.5^{+0.1}_{-0.6}$ \\
Abell 2147 & $4.56^{+0.1}_{-0.1}$ & $1.03^{+0.03}_{-0.009}$ & $3.37^{+0.07}_{-0.3}$ & $3.44^{+0.1}_{-0.08}$ \\
Abell 2163 & $23.4^{+1.}_{-1.}$ & $2.14^{+0.07}_{-0.1}$ & $35.1^{+4.}_{-6.}$ & $43.2^{+2.}_{-4.}$ \\
Abell 2204 & $9.47^{+0.3}_{-0.3}$ & $1.47^{+0.005}_{-0.06}$ & $11.0^{+0.1}_{-1.}$ & $13.6^{+0.2}_{-0.6}$ \\
Abell 2244 & $5.88^{+0.1}_{-0.1}$ & $1.18^{+0.02}_{-0.02}$ & $5.31^{+0.2}_{-0.3}$ & $6.04^{+0.2}_{-0.2}$ \\
Abell 2256 & $7.73^{+0.3}_{-0.3}$ & $1.37^{+0.03}_{-0.03}$ & $8.11^{+0.5}_{-0.6}$ & $9.54^{+0.8}_{-0.9}$ \\
Abell 2319 & $10.1^{+0.3}_{-0.2}$ & $1.58^{+0.01}_{-0.06}$ & $12.5^{+0.3}_{-1.}$ & $15.6^{+0.3}_{-0.7}$ \\
Abell 2657 & $3.82^{+0.1}_{-0.1}$ & $0.902^{+0.004}_{-0.03}$ & $2.28^{+0.04}_{-0.2}$ & $2.08^{+0.05}_{-0.07}$ \\
Abell 2734 & $4.28^{+0.2}_{-0.2}$ & $0.982^{+0.03}_{-0.007}$ & $3.00^{+0.05}_{-0.3}$ & $3.02^{+0.1}_{-0.08}$ \\
Abell 3112 & $5.01^{+2.}_{-1.}$ & $1.09^{+0.02}_{-0.02}$ & $4.19^{+0.2}_{-0.2}$ & $4.81^{+0.1}_{-0.1}$ \\
Abell 3158 & $5.20^{+0.06}_{-0.05}$ & $1.10^{+0.02}_{-0.01}$ & $4.23^{+0.2}_{-0.1}$ & $4.53^{+0.08}_{-0.05}$ \\
Abell 3376 & $4.58^{+0.1}_{-0.1}$ & $0.964^{+0.03}_{-0.002}$ & $2.80^{+0.03}_{-0.2}$ & $2.49^{+0.08}_{-0.02}$ \\
Abell 3391 & $5.41^{+0.4}_{-0.2}$ & $1.10^{+0.02}_{-0.02}$ & $4.19^{+0.3}_{-0.2}$ & $4.28^{+0.1}_{-0.1}$ \\
Abell 3571 & $7.73^{+0.3}_{-0.3}$ & $1.35^{+0.01}_{-0.03}$ & $7.60^{+0.2}_{-0.5}$ & $8.43^{+0.2}_{-0.2}$ \\
Abell 3667 & $6.60^{+0.01}_{-0.08}$ & $1.32^{+0.01}_{-0.03}$ & $7.33^{+0.2}_{-0.5}$ & $9.38^{+0.4}_{-0.3}$ \\
Abell 3822 & $5.36^{+0.3}_{-0.2}$ & $1.15^{+0.02}_{-0.02}$ & $4.82^{+0.2}_{-0.3}$ & $5.57^{+0.2}_{-0.2}$ \\
Abell 3827 & $7.89^{+0.2}_{-0.2}$ & $1.33^{+0.01}_{-0.04}$ & $7.60^{+0.3}_{-0.6}$ & $8.46^{+0.3}_{-0.4}$ \\
Abell 3921 & $5.93^{+0.2}_{-0.2}$ & $1.18^{+0.02}_{-0.02}$ & $5.39^{+0.2}_{-0.3}$ & $6.15^{+0.2}_{-0.1}$ \\
Abell 399 & $6.47^{+0.1}_{-0.1}$ & $1.26^{+0.01}_{-0.03}$ & $6.33^{+0.2}_{-0.4}$ & $7.43^{+0.1}_{-0.2}$ \\
Abell 400 & $2.15^{+0.05}_{-0.04}$ & $0.678^{+0.02}_{-0.04}$ & $0.955^{+0.07}_{-0.2}$ & $0.799^{+0.03}_{-0.07}$ \\
Abell 4038 & $3.12^{+0.05}_{-0.03}$ & $0.839^{+0.008}_{-0.03}$ & $1.82^{+0.05}_{-0.2}$ & $1.71^{+0.04}_{-0.09}$ \\
Abell 4059 & $4.34^{+0.1}_{-0.02}$ & $0.948^{+0.03}_{-0.004}$ & $2.66^{+0.2}_{-0.03}$ & $2.41^{+0.05}_{-0.07}$ \\
Abell 478 & $7.65^{+0.2}_{-0.2}$ & $1.39^{+0.006}_{-0.04}$ & $8.80^{+0.1}_{-0.7}$ & $11.3^{+0.4}_{-0.2}$ \\
Abell 539 & $2.59^{+0.04}_{-0.04}$ & $0.803^{+0.01}_{-0.03}$ & $1.59^{+0.05}_{-0.2}$ & $1.63^{+0.1}_{-0.03}$ \\
Abell 644 & $8.49^{+0.2}_{-0.1}$ & $1.30^{+0.02}_{-0.03}$ & $7.06^{+0.3}_{-0.5}$ & $6.86^{+0.3}_{-0.3}$ \\
Abell 754 & $11.8^{+0.7}_{-0.2}$ & $1.33^{+0.05}_{-0.04}$ & $7.44^{+0.8}_{-0.8}$ & $5.38^{+1.}_{-0.8}$ \\
Abell S 405 & $4.62^{+0.3}_{-0.3}$ & $0.985^{+0.03}_{-0.007}$ & $3.03^{+0.07}_{-0.3}$ & $2.85^{+0.09}_{-0.1}$ \\
Hydra A & $3.75^{+0.04}_{-0.03}$ & $1.02^{+0.03}_{-0.008}$ & $3.36^{+0.08}_{-0.2}$ & $4.20^{+0.07}_{-0.1}$ \\
Zw III 54 & $2.25^{+0.06}_{-0.06}$ & $0.718^{+0.01}_{-0.04}$ & $1.14^{+0.07}_{-0.2}$ & $1.05^{+0.03}_{-0.09}$ \\
MKW 3s & $3.44^{+0.09}_{-0.03}$ & $0.909^{+0.004}_{-0.03}$ & $2.35^{+0.03}_{-0.2}$ & $2.43^{+0.03}_{-0.1}$ \\
MKW 8 & $2.50^{+0.1}_{-0.1}$ & $0.716^{+0.02}_{-0.04}$ & $1.13^{+0.07}_{-0.2}$ & $0.924^{+0.05}_{-0.07}$ \\
PKS 0745-191 & $6.76^{+0.5}_{-0.2}$ & $1.35^{+0.01}_{-0.04}$ & $8.09^{+0.2}_{-0.6}$ & $11.1^{+0.4}_{-0.4}$ \\
UGC 3957 & $2.34^{+0.2}_{-0.1}$ & $0.697^{+0.02}_{-0.04}$ & $1.05^{+0.07}_{-0.2}$ & $0.864^{+0.05}_{-0.05}$ \\
ZwCl 1215+0400 & $7.57^{+0.3}_{-0.3}$ & $1.27^{+0.01}_{-0.03}$ & $6.57^{+0.2}_{-0.5}$ & $6.78^{+0.2}_{-0.2}$ \\
ZwCl 1742+3306 & $4.46^{+0.1}_{-0.1}$ & $0.957^{+0.03}_{-0.006}$ & $2.81^{+0.06}_{-0.2}$ & $2.61^{+0.07}_{-0.08}$ \\

\bottomrule
\end{tabular}
%\tablecomments{Global temperature, $kT_X$, the radius $R_{500}$, the total mass within $R_{500}$, $M_{500}$ and the gas mass within $R_{500}$, $M_{gas}$}
%\label{tab:}
\end{center}
\end{table}

%%%%%%%%%%%%%%% n(r) best-fit params %%%%%%%%%%%%%%%%%%
\clearpage
\section{Density and Temperature Profiles Best-Fit Parameters}
\label{app:profiles}

\begin{table}[here]
\begin{center}
\caption{Best-fit parameters of the electron density radial profiles.\label{tab:nofr}}
% updated to n_bestfit_params2_pr.tex on 2/16/2015
\begin{tabular}{ @{}lcccccc@{} }
\toprule
Cluster & $n_0$ & $\alpha$ & $\beta$ & $r_c$ & $r_s$ & $\epsilon$ \\ 
\midrule
Abell 119 & 0.0005406 & 0.627 & 5.0 & 2.959 & 0.289 & 1.227 \\ 
Abell 1413 & 0.04096 & 0.0 & 0.3753 & 0.0217 & 0.3968 & 2.13 \\ 
Abell 1644 & 0.04399 & 0.9245 & 0.3237 & 0.004989 & 2.17 & 5.0 \\ 
Abell 1651 & 0.009126 & 0.6828 & 0.3982 & 0.08472 & 0.2506 & 1.249 \\ 
Abell 1689 & 0.04991 & 0.0 & 0.399 & 0.0306 & 0.3455 & 1.99 \\ 
\bottomrule

\end{tabular}
\tablecomments{Table \ref{tab:nofr} is published in its entirety in the electronic edition of the Astrophysical Journal. A portion is shown here for guidance regarding its form and content. The $n_e(r)$ model is shown in Equation \ref{eq:vikhnofr}.}
\end{center}
\end{table}

%%%%%%%%%%%%%%% kT best-fit params %%%%%%%%%%%%%%%%%%
%\clearpage
%\section{Temperature Profile Best-fit Parameters}

\begin{table}[here]
\begin{center}
\caption{Best-fit parameters of the temperature radial profiles.\label{tab:ktofr}} 
% updated to kt_bestfit_params2_pr.tex on 2/16/2015
\begin{tabular}{ @{}lcccccccc@{} }
\toprule
Cluster & $r_t$ & $a$ & $b$ & $c$ & $a_{cool}$ & $r_{cool}$ & $t_{min}$ & $t_0$ \\ 
\midrule
Abell 119 & 0.2912 & -0.3165 & 4.758 & 1.011 & 12.51 & 0.08264 & 19.22 & 8.419 \\ 
Abell 1413 & 0.01754 & 0.4569 & 6.889 & -0.7317 & -0.7636 & 0.1653 & 1.329 & 7.881 \\ 
Abell 1644 & 0.4703 & -0.2443 & 7.74 & 0.5485 & 3.737 & 0.02752 & 4.614 & 6.027 \\ 
Abell 1651 & 0.04094 & 0.04568 & 4.947 & -0.0793 & 1.9 & 0.05 & 6.693 & 6.858 \\ 
Abell 1689 & 1.046 & -0.002614 & 7.542 & 3.02 & 8.231 & 0.07014 & 10.04 & 11.45 \\ 
\bottomrule
\end{tabular}
\tablecomments{Table \ref{tab:ktofr} is published in its entirety in the electronic edition of the Astrophysical Journal. A portion is shown here for guidance regarding its form and content. The $kT(r)$ model is shown in Equation \ref{eq:ktofr}.}
\end{center}
\end{table}

%%%%%%%%%%%%%%% csb, s40, <w> %%%%%%%%%%%%%%%%%%
%\clearpage
\section{Morphological Parameters}

\begin{table}[h]
\begin{center}
\caption{Morphological Parameters and Entropy near the Center.\label{tab:morph}}
% updated to table_morph_ac3b_pr.tex results 2/16/2015
\begin{tabular}{ @{}llll@{} }
\toprule
Cluster & $c_{SB}$\tablenotemark{a} & $S_{40}$\tablenotemark{b}(keV cm$^2$)  & $\langle w \rangle\tablenotemark{c}/10^{-3}$ \\
\midrule
Abell 119 & 2.17$\pm$ 0.037 & $550.^{+69.}_{-65.}$ & 1.12 \\
Abell 1413 & 9.91$\pm$ 0.079 & $112.^{+4.0}_{-3.7}$ & 0.607 \\
Abell 1644 & 6.13$\pm$ 0.057 & $95.4^{+2.6}_{-2.1}$ & (4.60) \\
Abell 1651 & 7.71$\pm$ 0.16 & $140.^{+4.2}_{-4.0}$ & 3.15 \\
Abell 1689 & 12.3$\pm$ 0.064 & $108.^{+2.8}_{-2.4}$ & 1.16 \\
\bottomrule
\end{tabular}
\tablenotetext{1}{Surface brightness concentration.}
\tablenotetext{2}{Entropy at $r=$40 kpc.}
\tablenotetext{3}{Centroid shift.}
\tablecomments{Table \ref{tab:morph} is published in its entirety in the electronic edition of the Astrophysical Journal. A portion is shown here for guidance regarding its form and content. Values of $\langle w \rangle$ in parentheses correspond to measurements in observations where the FOV does not fully cover the region $r<0.3R_{500}$, and are excluded from analysis involving $\langle w \rangle$, but shown here.}
%\label{tab:}
\end{center}
\end{table}

%%%%%%%%%%%%%%% Zmid and Zin %%%%%%%%%%%%%%%%%%

\clearpage
\section{Global Metallicity Measures}
\label{app:globalz}

\begin{table}[h]
\begin{center}
\caption{Global metallicity measures.\label{tab:globalz}}
\begin{tabular}{ @{}llll@{} }
% updated to values in table_abund_ac3_pr.tex on 2/16/2015
\toprule
Cluster & $\bar{Z}_{mid}(Z_\odot)$ & $\bar{Z}_{in}(Z_\odot)$  \\
\midrule
Abell 119 & 0.260$^{+0.036}_{-0.035}$ & 0.337$^{+0.047}_{-0.045}$ \\
Abell 1413 & 0.253$^{+0.032}_{-0.032}$ & 0.372$^{+0.027}_{-0.026}$ \\
Abell 1644 & 0.226$^{+0.029}_{-0.028}$ & 0.360$^{+0.028}_{-0.026}$ \\
Abell 1651 & 0.348$^{+0.081}_{-0.076}$ & 0.385$^{+0.069}_{-0.064}$ \\
Abell 1689 & 0.304$^{+0.032}_{-0.031}$ & 0.352$^{+0.025}_{-0.024}$ \\
Abell 1736 & 0.283$^{+0.036}_{-0.034}$ & 0.306$^{+0.046}_{-0.044}$ \\
Abell 1795 & 0.240$^{+0.023}_{-0.023}$ & 0.324$^{+0.013}_{-0.013}$ \\
Abell 1914 & 0.314$^{+0.095}_{-0.091}$ & 0.404$^{+0.064}_{-0.061}$ \\
Abell 2029 & 0.238$^{+0.024}_{-0.023}$ & 0.421$^{+0.015}_{-0.015}$ \\
Abell 2063 & 0.187$^{+0.024}_{-0.022}$ & 0.327$^{+0.022}_{-0.021}$ \\
Abell 2065 & 0.221$^{+0.043}_{-0.040}$ & 0.316$^{+0.041}_{-0.039}$ \\
Abell 2142 & 0.366$^{+0.037}_{-0.036}$ & 0.328$^{+0.023}_{-0.022}$ \\
Abell 2147 & 0.290$^{+0.038}_{-0.036}$ & 0.269$^{+0.039}_{-0.037}$ \\
Abell 2163 & 0.476$^{+0.10}_{-0.10}$ & 0.345$^{+0.077}_{-0.075}$ \\
Abell 2204 & 0.397$^{+0.043}_{-0.042}$ & 0.391$^{+0.025}_{-0.024}$ \\
Abell 2244 & 0.201$^{+0.026}_{-0.025}$ & 0.298$^{+0.021}_{-0.020}$ \\
Abell 2256 & 0.289$^{+0.086}_{-0.078}$ & 0.389$^{+0.065}_{-0.061}$ \\
Abell 2319 & 0.286$^{+0.052}_{-0.050}$ & 0.377$^{+0.049}_{-0.047}$ \\
Abell 2657 & 0.281$^{+0.043}_{-0.040}$ & 0.460$^{+0.055}_{-0.051}$ \\
Abell 2734 & 0.227$^{+0.056}_{-0.052}$ & 0.227$^{+0.056}_{-0.052}$ \\
Abell 3112 & 0.182$^{+0.034}_{-0.032}$ & 0.394$^{+0.029}_{-0.027}$ \\
Abell 3158 & 0.389$^{+0.026}_{-0.024}$ & 0.420$^{+0.029}_{-0.027}$ \\
Abell 3376 & 0.312$^{+0.042}_{-0.040}$ & 0.225$^{+0.044}_{-0.042}$ \\
Abell 3391 & 0.228$^{+0.069}_{-0.064}$ & 0.228$^{+0.069}_{-0.064}$ \\
Abell 3571 & 0.191$^{+0.034}_{-0.032}$ & 0.392$^{+0.024}_{-0.023}$ \\
Abell 3667 & 0.309$^{+0.010}_{-0.010}$ & 0.345$^{+0.012}_{-0.011}$ \\
Abell 3822 & 0.258$^{+0.067}_{-0.063}$ & 0.258$^{+0.067}_{-0.063}$ \\
Abell 3827 & 0.265$^{+0.052}_{-0.049}$ & 0.368$^{+0.038}_{-0.036}$ \\
Abell 3921 & 0.327$^{+0.051}_{-0.049}$ & 0.308$^{+0.041}_{-0.039}$ \\
Abell 399 & 0.283$^{+0.047}_{-0.045}$ & 0.307$^{+0.049}_{-0.046}$ \\
Abell 400 & 0.394$^{+0.040}_{-0.036}$ & 0.548$^{+0.059}_{-0.053}$ \\
Abell 4038 & 0.355$^{+0.025}_{-0.023}$ & 0.440$^{+0.020}_{-0.019}$ \\
Abell 4059 & 0.250$^{+0.017}_{-0.016}$ & 0.466$^{+0.013}_{-0.012}$ \\
Abell 478 & 0.257$^{+0.025}_{-0.024}$ & 0.328$^{+0.016}_{-0.015}$ \\
Abell 539 & 0.212$^{+0.023}_{-0.021}$ & 0.225$^{+0.024}_{-0.022}$ \\
Abell 644 & 0.323$^{+0.044}_{-0.042}$ & 0.438$^{+0.038}_{-0.036}$ \\
Abell 754 & 0.302$^{+0.058}_{-0.057}$ & 0.369$^{+0.036}_{-0.035}$ \\
Abell S 405 & 0.264$^{+0.083}_{-0.077}$ & 0.264$^{+0.083}_{-0.077}$ \\
Hydra A & 0.147$^{+0.0095}_{-0.0094}$ & 0.236$^{+0.0055}_{-0.0054}$ \\
Zw III 54 & 0.258$^{+0.029}_{-0.026}$ & 0.370$^{+0.037}_{-0.034}$ \\
MKW 3s & 0.328$^{+0.025}_{-0.024}$ & 0.395$^{+0.017}_{-0.017}$ \\
MKW 8 & 0.241$^{+0.045}_{-0.039}$ & 0.466$^{+0.082}_{-0.075}$ \\
PKS 0745-191 & 0.206$^{+0.045}_{-0.043}$ & 0.369$^{+0.028}_{-0.026}$ \\
UGC 3957 & 0.215$^{+0.051}_{-0.044}$ & 0.426$^{+0.049}_{-0.043}$ \\
ZwCl 1215+0400 & 0.315$^{+0.060}_{-0.056}$ & 0.357$^{+0.069}_{-0.065}$ \\
ZwCl 1742+3306 & 0.314$^{+0.045}_{-0.042}$ & 0.447$^{+0.026}_{-0.025}$ \\
\bottomrule
\end{tabular}
\end{center}
\end{table}

\end{document}